\documentclass{article}

\PassOptionsToPackage{numbers,compress}{natbib}

 \usepackage[main, final]{neurips_2026}

\usepackage[utf8]{inputenc} 
\usepackage[T1]{fontenc}    
\usepackage{hyperref}       
\usepackage{url}            
\usepackage{booktabs}       
\usepackage{pifont}         
\usepackage{amsfonts}       
\usepackage{amsmath}        
\usepackage{amssymb}        
\usepackage{nicefrac}       
\usepackage{microtype}      
\usepackage{xcolor}         
\usepackage[normalem]{ulem} 
\usepackage{graphicx}       
\usepackage{algorithm}      
\usepackage{algpseudocode}  
\usepackage{wrapfig}        
\usepackage{float}          
\usepackage{tikz}           

\title{Velocity-Space 3D Asset Editing}

\author{%
  Hao Liu$^{1,3}$\quad
  Yuxuan Lin$^{1}$\quad
  Jingfeng Guo$^{2}$\quad
  Ruihang Chu$^{1}$\quad
  Junjie Wang$^{1}$ \\
  \setcounter{footnote}{1}%
  \textbf{Ruotong Li}$^{3}$\footnotemark[2]\quad
  \textbf{Yujiu Yang}$^{1}$\thanks{Corresponding authors.} \\[0.5em]
  $^{1}$Tsinghua University \quad
  $^{2}$South China University of Technology \quad
  $^{3}$\mbox{Pengcheng Laboratory} \\[0.3em]
  {\small\texttt{\{h-l25, rhchu\}@mails.tsinghua.edu.cn, linyx23@tsinghua.org.cn}} \\
  {\small\texttt{drjfguo@gmail.com, \{wangjunjie, yang.yujiu\}@sz.tsinghua.edu.cn, lirt\_cs@163.com}}
}

\begin{document}

\maketitle

\begin{abstract}
Editing a 3D asset locally, modifying a target region while preserving the rest, is a fundamental requirement of native 3D editing. Existing methods enforce locality through mechanisms external to the generator, such as manual 3D masks, post-hoc voxel merging, or 2D multi-view lifting. None of them intervene where the corruption actually originates: inside the ODE sampler. For a rectified-flow generator to achieve faithful local editing, its velocity field should be strong over the target editing region while vanishing on preserved content. Yet a single velocity field can hardly satisfy both requirements simultaneously, leading to three problems: (i) identity leakage that keeps the edit signal non-zero on preserved regions; (ii) no dedicated edit-amplification channel, so strengthening the edit inevitably perturbs identity; and (iii) an identity drag at the geometry and material stages, where a global condition pulls every token toward the target.  \textcolor{black}{We propose \textbf{VS3D} (\emph{\textbf{V}elocity-\textbf{S}pace \textbf{3D} Asset editing}), an \textbf{inversion-free, training-free, and mask-free} framework that derives edit and preserve signals by comparing the frozen model's source- and target-conditioned responses and realizes this principle through three stage-specific modules.} Reconstruction-Anchored Source Injection (RASI) absorbs identity leakage by turning the unconditional embedding into a per-step, asset-specific anchor calibrated through source reconstruction. Partial-Mean Guidance (PMG) amplifies the edit signal by contrasting high- and low-quality subsample estimates of the velocity difference, active only where a consistent edit exists. \textcolor{black}{Twin-Agreement Residual Injection} (TAR) lets the sampler decide token by token what to preserve at the geometry and material stages.  \textcolor{black}{Experiments on diverse 3D assets show that VS3D outperforms state-of-the-art native-3D and 2D-lifted editors, demonstrating the effectiveness of velocity-space editing for training-free 3D asset editing. Code is publicly available at \url{https://github.com/LHaoooo/VS3D}.}
\end{abstract}


\section{Introduction}
\label{sec:intro}

Precise local editing of 3D assets is essential for content creation, game development, and embodied simulation, yet it poses a fundamental dual requirement on any pretrained generator: the model must apply a strong, targeted modification to the region the user specifies while leaving every other region strictly unchanged. The dominant family of native 3D generators~\citep{trellis2,hunyuan3d25,sparc3d,direct3ds2,lattice} now builds on rectified-flow DiTs~\citep{liu2022flow,lipman2023flow}, where generation is an ODE whose velocity field is learned by conditional flow matching. Editing with such a model therefore reduces to \emph{controlling its velocity field}: the velocity update must carry a non-trivial edit signal on the target region and vanish on the rest. How to achieve this control on a frozen 3D rectified-flow DiT, without resorting to external masks or additional training, remains an open problem.

\textcolor{black}{Existing 3D editors typically enforce source preservation through mechanisms outside the velocity-field update.} Three routes recur: (i) \emph{mask-based editors}~\citep{voxhammer,easy3e,vecsetedit,vinedresser3d} ask the user or an external tool to supply a binary 3D region that pins every non-edited voxel to the source, tedious to draw on a voxel grid, brittle for natural edits whose support fades gradually, and prone to seam artefacts wherever a hard binary boundary meets smooth geometry; (ii) \emph{post-hoc mergers}~\citep{nano3d} try to recover identity by connected-component stitching that can neither distinguish genuine edits from leakage nor suppress boundary artefacts at the merge seam; and (iii) \emph{multi-view / video editors}~\citep{edit360,instant3dit} enforce identity in 2D and lift the result back through a reconstructor, yet edits that alter topology or volumetric occupancy introduce multi-view inconsistencies that accumulate across views. All three share a deeper limitation: because \textbf{the velocity field of a rectified-flow ODE is the sole carrier of both identity and edit signals}, the most direct path to simultaneous identity preservation and edit amplification is to control that velocity field from within; none of the existing routes attempt this, and instead impose identity from outside the ODE without ever examining the ODE itself.

Among inversion-free velocity-space editing formulations, FlowEdit~\citep{flowedit} couples a source and a target branch through shared noise so that their differential velocity $v_{\Delta}$ directly drives the edit offset without inverting the source into noise. We adopt this coupling and \emph{examine the ODE itself} to understand why identity still drifts. Our analysis reveals three problems. \textbf{(P1) Two-channel identity leakage.} Even when the two conditions locally agree, $v_{\Delta}$ does not vanish on preserved content: it carries a \emph{CFG-asymmetry residual}, structurally non-zero because effective editing requires a much stronger target guidance than source guidance, and an \emph{attention-mediated condition residual}, because \textcolor{black}{the conditioning features are a sequence of tokens extracted from a single image, whose perturbations can propagate through cross-attention to voxel tokens outside the edited region}. Euler integration accumulates both residuals and visibly deforms non-edited regions (closed-form derivation in \S\ref{sec:rasi}). \textbf{(P2) Absence of a dedicated edit-amplification channel.} Identity preservation and edit amplification compete for the same degrees of freedom within the velocity field. Because few-sample inference is unavoidable for costly 3D DiTs, the Monte-Carlo estimate of $v_{\Delta}$ is dominated by estimation variance rather than the consistent edit signal, leaving the edit region under-expressed even when the leakage has been suppressed. \textbf{(P3) Global-condition override.} The downstream geometry and material stages refine shape and appearance on the coordinate occupancy produced by the sparse-occupancy stage, \textcolor{black}{each conditioned on feature tokens extracted from a single \emph{target} image, which can pull non-edited tokens toward the target appearance}. Without a counterbalancing mechanism, this global drag overwrites the source identity on preserved tokens, causing geometry distortion and texture drift. \textcolor{black}{VS3D addresses these failures inside the sampler at the stage where each problem arises.}

 \textcolor{black}{We address each problem with a targeted intervention inside the ODE, and the three solutions share a common insight: edit-sensitive content is identified by disagreement between the frozen model's source- and target-conditioned responses, while agreement indicates content to preserve. Because this signal appears differently in dense occupancy latents and sparse geometry/material SLATs, VS3D realizes the same criterion with stage-specific mechanisms.} For \textbf{P1}, Reconstruction-Anchored Source Injection (RASI) turns the unconditional embedding into a per-step, asset-specific handle optimised through a source-reconstruction probe, absorbing both leakage channels at once so that $v_{\Delta}$ collapses toward zero on non-edited voxels without any external region indication. For \textbf{P2}, Partial-Mean Guidance (PMG) contrasts a high-quality and a low-quality subsample estimate of the edit direction and extrapolates away from the noisier one; the amplification is automatically gated by $v_{\Delta}$ itself, active only where a consistent edit signal survives and inert where RASI has already suppressed the velocity. For \textbf{P3}, \textcolor{black}{Twin-Agreement Residual Injection} (TAR) runs a second sparse-DiT pass on the same coordinate occupancy, but conditioned on the source image, and measures per-token agreement; tokens on which the two passes agree are blended back toward the source encoding, while the edit region is left untouched. We refer to the resulting system as \textbf{VS3D}, an inversion-free, training-free, mask-free native-3D editor that achieves high-fidelity local edits with strong identity preservation entirely inside the ODE, \textcolor{black}{without relying on external mechanisms to enforce edit locality.}

\paragraph{Contributions.} Our main contributions are as follows:
{\color{black}
\begin{itemize}\setlength{\itemsep}{0pt}\setlength{\parskip}{0pt}
  \item We formulate local 3D editing as a model-internal edit/preserve problem in velocity space, where disagreement between source- and target-conditioned responses indicates edit-sensitive content and agreement indicates content to preserve.
  \item We implement this criterion with stage-specific modules: Reconstruction-Anchored Source Injection (RASI; \S\ref{sec:rasi}) and Partial-Mean Guidance (PMG; \S\ref{sec:pmg}) act on dense occupancy latents to suppress leakage and amplify edits, while \textcolor{black}{Twin-Agreement Residual Injection} (TAR; \S\ref{sec:tar}) acts on sparse geometry and material SLATs to preserve non-edited tokens.
  \item We build \textbf{VS3D}, an inversion-free, training-free, and mask-free full-asset editing framework that edits occupancy, geometry, and material within a frozen 3D ODE sampler.
\end{itemize}}

\section{Background}
\label{sec:background}

\paragraph{Native structured-latent 3D generator.}
We adopt TRELLIS~2.0~\citep{trellis2} as a frozen backbone: a native 3D generator whose generation process unfolds in three stages with three independent DiTs, all conditioned on a DINOv3-L~\citep{dinov3} embedding of the input image. The three stages are (i) \emph{sparse-structure generation}, (ii) \emph{geometry generation}, and (iii) \emph{material generation}. Stage~1 operates on a \emph{dense} low-resolution latent $z^{\mathrm{ss}}\!\in\!\mathbb{R}^{C\times R\times R\times R}$ ($R{=}64$ in our $1024^{3}$-output setting) and is decoded into a binary occupancy grid whose active coordinates $\mathcal{V}\!\subset\!\mathbb{Z}^{3}$ define the voxel support of the subsequent stages. \textcolor{black}{Stages~2--3 use sparse DiTs on $\mathcal{V}$, with separate structured latents (SLATs) for geometry and material. Two Sparse Compression VAEs (SC-VAEs), each with $16{\times}$ spatial compression, decode these latents into geometry and PBR material fields (albedo, roughness, metallic, and normal). During upsampling, the material SC-VAE is conditioned on the geometry decoder's subdivision structure~\citep{trellis2}, aligning the geometry and PBR fields on the shared O-Voxel support.}

\paragraph{Rectified flow.}
Each stage of TRELLIS~2.0 is a rectified flow~\citep{liu2022flow} trained under the conditional flow-matching objective~\citep{lipman2023flow}. For a clean latent $x_{0}$ and Gaussian noise $\varepsilon\!\sim\!\mathcal{N}(0,\mathbf{I})$, the forward process is the straight-line interpolation
\begin{equation}
x(t) \;=\; (1-t)\,x_{0} \;+\; t\,\varepsilon,\qquad t\in[0,1],
\label{eq:rf-interp}
\end{equation}
whose constant time derivative $\varepsilon-x_{0}$ is the regression target of the velocity network $v_{\theta}(x,t,c)$, where $c$ denotes the image-conditioning signal. Setting $v_{\theta}(x,t,c)\!\approx\!\varepsilon-x_{0}$ and eliminating $\varepsilon$ via~\eqref{eq:rf-interp} gives the implicit clean-sample estimate
\begin{equation}
\hat x_{0}(x,t,c) \;=\; x - t\,v_{\theta}(x,t,c).
\label{eq:x0-estimate}
\end{equation}

\paragraph{Inversion-free velocity-space editing.}
Given the clean source latent $x_{\mathrm{src}}$ and conditions $c_{\mathrm{src}},c_{\mathrm{tgt}}$, FlowEdit~\citep{flowedit} transports $x_{\mathrm{src}}$ to an edited latent without ever inverting it into a noise latent. It maintains an edit state $z^{\mathrm{edit}}_{t}$ initialised at $z^{\mathrm{edit}}_{1}{=}x_{\mathrm{src}}$ and integrated from $t{=}1$ down to $t{=}0$ along a discrete schedule $1{=}t_{0}{>}t_{1}{>}\cdots{>}t_{T}{=}0$ with step size $\Delta t{=}t_{k+1}-t_{k}$. At each step the two branches are coupled by a shared noise draw $\varepsilon\!\sim\!\mathcal{N}(0,\mathbf{I})$ on the interpolation of~\eqref{eq:rf-interp},
\begin{equation}
z_{t}^{\mathrm{src}} = (1-t)\,x_{\mathrm{src}} + t\,\varepsilon, \qquad
z_{t}^{\mathrm{tgt}} = z_{t}^{\mathrm{edit}} + \big(z_{t}^{\mathrm{src}} - x_{\mathrm{src}}\big),
\label{eq:flowedit-coupling}
\end{equation}
so that $z_{t}^{\mathrm{tgt}}-z_{t}^{\mathrm{src}}$ equals the running edit offset $z_{t}^{\mathrm{edit}}-x_{\mathrm{src}}$ at every $t$. Each branch is then sampled under its own condition and its own classifier-free guidance weight,
\begin{equation}
\tilde v_{\theta}(z_{t},t,c,\phi;\omega) \;=\; (1+\omega)\,v_{\theta}(z_{t},t,c) \;-\; \omega\,v_{\theta}(z_{t},t,\phi),
\label{eq:cfg}
\end{equation}
where $\omega$ is the guidance weight and $\phi$ is an unconditional embedding (by default the network's built-in null $\phi_{0}$). Since $z_{t}^{\mathrm{tgt}}-z_{t}^{\mathrm{src}}$ tracks the edit offset itself, the direction along which this offset should evolve is exactly the difference of the two branches' guided velocities:
\begin{equation}
v_{\Delta}(z^{\mathrm{edit}}_{t},t) \;=\; \tilde v_{\theta}(z_{t}^{\mathrm{tgt}},t,c_{\mathrm{tgt}},\phi;\omega_{\mathrm{tgt}}) \;-\; \tilde v_{\theta}(z_{t}^{\mathrm{src}},t,c_{\mathrm{src}},\phi;\omega_{\mathrm{src}}).
\label{eq:vdelta}
\end{equation}
The edit state is advanced by one Euler step $z^{\mathrm{edit}}_{t_{k+1}} = z^{\mathrm{edit}}_{t_{k}} + \Delta t\,v_{\Delta}(z^{\mathrm{edit}}_{t_{k}},t_{k})$, where $v_{\Delta}$ is estimated as a \textbf{Monte-Carlo average over $S$ independent noise realisations} of the coupling~\eqref{eq:flowedit-coupling}, with $S$ kept small on a 3D rectified-flow DiT due to per-step cost. Integrating $v_{\Delta}$ thus transports $x_{\mathrm{src}}$ to an edited latent directly in velocity space, without ever inverting the source into a noise latent.

\section{VS3D}
\label{sec:method}

\subsection{Overview}
\label{sec:overview}

\begin{figure}[t]
  \centering
  \includegraphics[width=\textwidth]{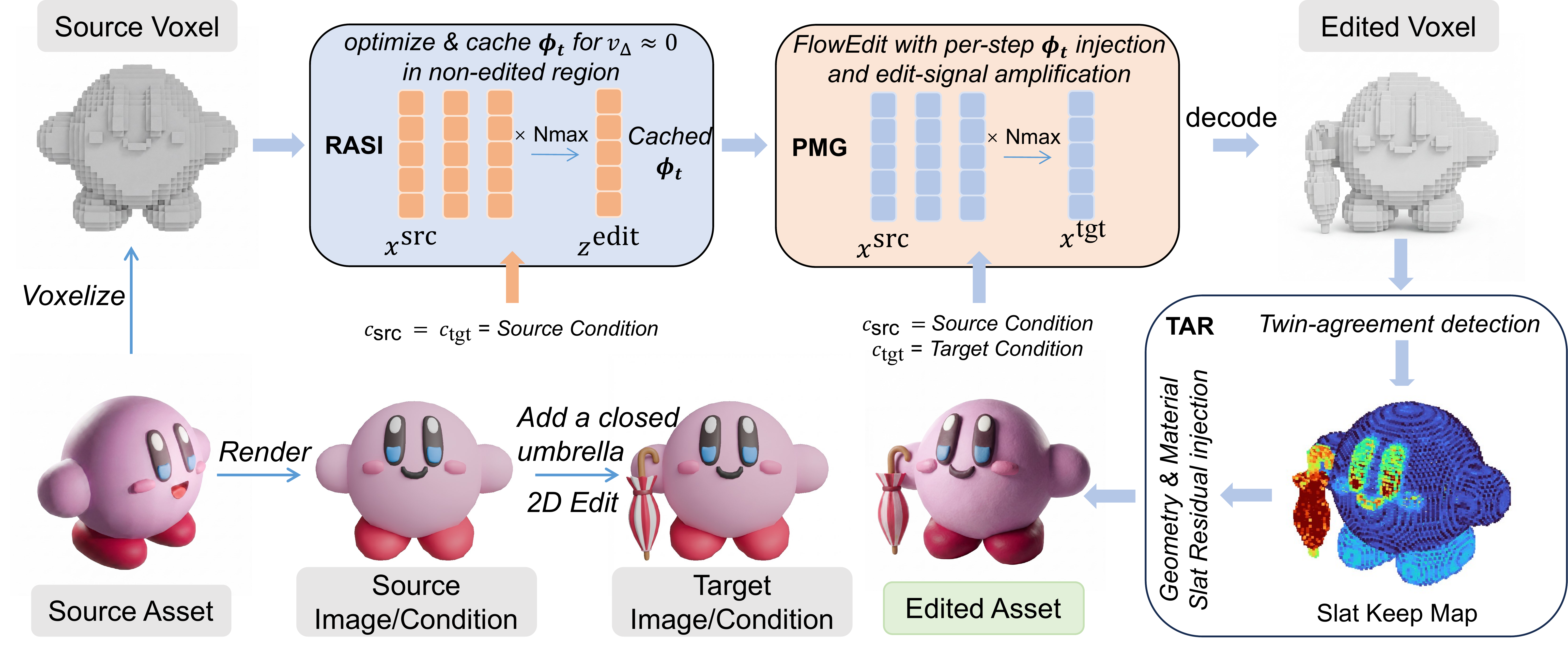}
  \caption{\textbf{Overview of the VS3D pipeline.} \textcolor{black}{The source 3D asset is rendered to obtain a source image, which is edited in 2D to produce the target condition image.} Stage~1 operates on the dense occupancy latent: \emph{RASI} (\S\ref{sec:rasi}) optimises a per-step $\phi_t$ to suppress $v_{\Delta}$ on non-edited regions, and \emph{PMG} (\S\ref{sec:pmg}) amplifies the edit signal via subsample extrapolation. Stages~2--3 handle sparse geometry and material SLATs: \emph{TAR} (\S\ref{sec:tar}) computes a token-wise $p_{\mathrm{keep}}$ map (blue\,=\,preserve, red\,=\,edit) and injects source residuals accordingly to produce the final edited asset.}
  \label{fig:pipeline}
\end{figure}

VS3D addresses mask-free local editing with three modules acting on the ODE, each targeting a distinct failure mode (Fig.~\ref{fig:pipeline}). The editing problem naturally splits into two regimes. The \emph{dense} Stage-1 occupancy latent admits FlowEdit's coupling~\eqref{eq:flowedit-coupling}, so editing reduces to shaping $v_{\Delta}$; however, two leakage channels---CFG asymmetry and the global-condition gap (\S\ref{sec:rasi})---keep $v_{\Delta}$ structurally non-zero on preserved content. The \emph{sparse} Stage-2/3 geometry and material SLATs have mismatched active coordinates between source and target, making velocity-space coupling unavailable; \textcolor{black}{DINOv3 feature tokens extracted from a single image can spread changes in the target condition to non-edited tokens}, so the challenge becomes \emph{which tokens to leave alone} (\S\ref{sec:tar}).

Our three modules resolve these issues in sequence:
\textbf{(i)}~\emph{Reconstruction-Anchored Source Injection} (RASI, \S\ref{sec:rasi}) closes Stage-1 identity leakage by turning the null embedding into a per-step, asset-specific $\phi_{t_{k}}$ fitted via a source-reconstruction probe, absorbing both leakage channels simultaneously.
\textbf{(ii)}~\emph{Partial-Mean Guidance} (PMG, \S\ref{sec:pmg}) amplifies the edit signal by extrapolating from a noisier subsample mean of $v_{\Delta}$ toward a cleaner one; the amplification is automatically gated by $v_{\Delta}$ itself, active only where a consistent edit exists.
\textbf{(iii)}~\emph{Twin-Agreement Residual Injection} (TAR, \S\ref{sec:tar}) lets the Stage-2/3 sparse-DiT sampler decide token by token what to preserve and what to regenerate, handling additions, removals, and replacements uniformly.

\paragraph{Problem formulation.}
Given a source 3D asset $\mathcal{A}_{\mathrm{src}}$ with geometry and material, and a pair of reference images $(I_{\mathrm{src}}, I_{\mathrm{tgt}})$ depicting its current and desired appearance, we seek an edited asset that realises the edit in the implied region while leaving the rest intact, under two hard constraints:
\begin{itemize}\setlength{\itemsep}{0pt}\setlength{\parskip}{0pt}
  \item \textbf{Training-free}: no model or LoRA is fine-tuned; all parameters of $v_{\theta}$ are frozen.
  \item \textbf{Mask-free}: neither manual masks nor external segmentation networks are used to delineate the edit region.
\end{itemize}

\subsection{Reconstruction-Anchored Source Injection for Identity-Leakage Closure}
\label{sec:rasi}

The velocity difference $v_{\Delta}$ should vanish on non-edited voxels, yet off-the-shelf FlowEdit leaves a persistent residual that Euler integration accumulates into identity drift. We trace this residual to two structural channels and close both by turning the unconditional embedding $\phi$ into a per-step, asset-specific calibration \textcolor{black}{variable} optimised through a source-reconstruction probe.

\paragraph{Problem diagnosis.}
To isolate the first leakage channel, consider forcing $c_{\mathrm{tgt}}=c_{\mathrm{src}}$: the two branches of~\eqref{eq:flowedit-coupling} then share the same condition and should ideally give $v_{\Delta}\!=\!0$, yet substituting $c_{\mathrm{tgt}}=c_{\mathrm{src}}$ into~\eqref{eq:vdelta} and expanding~\eqref{eq:cfg} yields
\begin{equation}
v_{\Delta}\big|_{c_{\mathrm{tgt}}=c_{\mathrm{src}}} \;=\; (\omega_{\mathrm{tgt}}-\omega_{\mathrm{src}})\,\big[\,v_{\theta}(z_{t}^{\mathrm{src}},t,c_{\mathrm{src}}) \,-\, v_{\theta}(z_{t}^{\mathrm{src}},t,\phi)\,\big].
\label{eq:vdelta-residual}
\end{equation}
Effective editing adopts $\omega_{\mathrm{tgt}}\!\gg\!\omega_{\mathrm{src}}$ for the target condition to take effect, so the prefactor in~\eqref{eq:vdelta-residual} is structurally non-zero even in this degenerate regime: asymmetric CFG weights alone are already enough to keep $v_{\Delta}$ away from zero, independent of any condition gap. Second, local agreement on a non-edited region does not imply global agreement: \textcolor{black}{since DINOv3 encodes each image into a sequence of contextualized feature tokens}, any difference in the edited region perturbs $c_{\mathrm{src}},c_{\mathrm{tgt}}$ as a whole, and dense cross-attention then propagates this gap into every voxel token, so $v_{\theta}(\cdot,t,c_{\mathrm{src}})\neq v_{\theta}(\cdot,t,c_{\mathrm{tgt}})$ even on non-edited voxels. Euler integration accumulates both residuals into $z_{t}^{\mathrm{edit}}$ along the non-edited support, and mask-free editing cannot remove them after the fact.

\paragraph{Absorbing leakage via $\phi$.}
Among the symbols in~\eqref{eq:vdelta-residual}, $v_{\theta}$ is frozen, $c_{\mathrm{src}},c_{\mathrm{tgt}}$ carry the edit semantics, and $\omega_{\mathrm{tgt}},\omega_{\mathrm{src}}$ set the edit strength. \textbf{The unconditional embedding $\phi$ is the natural absorption site for both leakage channels}: optimising $\phi$ per timestep encodes asset-specific reconstruction information without interfering with the condition or the guidance schedule~\citep{nti}. We design a source-reconstruction probe: at each active step $t_{k}$, we replace $c_{\mathrm{tgt}}$ with $c_{\mathrm{src}}$ on the target branch while keeping its guidance weight at $\omega_{\mathrm{tgt}}$, and ask whether a single Euler step of $v_{\Delta}$ can bring $z_{t_{k}}^{\mathrm{edit}}$ back to $x_{\mathrm{src}}$. Switching only the condition but not the CFG weight reproduces the exact guidance schedule used at editing time, so both leakage channels, including the $(\omega_{\mathrm{tgt}}-\omega_{\mathrm{src}})$ prefactor, are present in the probe and available to be absorbed by $\phi$. Concretely, we solve
\begin{equation}
\phi_{t_{k}} \;=\; \arg\min_{\phi}\; \Big\| \,z_{t_{k}}^{\mathrm{edit}} + \Delta t \big(\tilde v_{\theta}(z_{t_{k}}^{\mathrm{tgt}},t_{k},c_{\mathrm{src}},\phi;\omega_{\mathrm{tgt}}) - \tilde v_{\theta}(z_{t_{k}}^{\mathrm{src}},t_{k},c_{\mathrm{src}},\phi;\omega_{\mathrm{src}})\big) \,-\, x_{\mathrm{src}} \Big\|_{2}^{2},
\label{eq:rasi-obj}
\end{equation}
where $z_{t_{k}}^{\mathrm{src}},z_{t_{k}}^{\mathrm{tgt}}$ are formed via the coupling~\eqref{eq:flowedit-coupling}, \emph{both} branches are conditioned on $c_{\mathrm{src}}$, and each branch retains its real-editing guidance weight. Minimising the deviation from $x_{\mathrm{src}}$ forces $\phi_{t_{k}}$ to absorb both leakage channels into a single correction tied to this asset and this timestep. We then advance $z^{\mathrm{edit}}$ one Euler step along the probe ODE and cache $\phi_{t_{k}}$; optimisation hyper-parameters are reported in~\S\ref{app:setup}.

During editing (\S\ref{sec:pmg}), we evaluate $v_{\Delta}$ with $c_{\mathrm{tgt}}$ on the target branch and $c_{\mathrm{src}}$ on the source branch, substituting the cached $\phi_{t}$ for $\phi_{0}$ in both CFG calls. Because the probe has pinned the $c_{\mathrm{src}}$-only behaviour of $v_{\Delta}$ to self-reconstruction, the real update retains only the $c_{\mathrm{tgt}}$-driven edit signal: on non-edited voxels $v_{\Delta}$ collapses toward zero without any mask, turning identity preservation into a property of the velocity field itself.

\subsection{Partial-Mean Guidance for Edit-Signal Amplification}
\label{sec:pmg}

RASI secures identity on non-edited voxels, but the same optimisation may inadvertently suppress $v_{\Delta}$ on edited ones as well, weakening the edit signal. We need an amplification handle that is automatically gated by the edit region---active where $v_{\Delta}$ is large, silent where RASI has driven it toward zero---so that edit strength and identity preservation are decoupled without any mask. We obtain such a handle from the Monte-Carlo structure of $v_{\Delta}$ itself, at no extra forward cost, by extrapolating between a full-sample mean and a noisier partial-sample mean.

\paragraph{Why a separate amplifier is needed.}
Because the RASI objective~\eqref{eq:rasi-obj} asks one Euler step of $v_{\Delta}$ to reconstruct $x_{\mathrm{src}}$, the optimised $\phi_{t}$ may absorb $v_{\Delta}$ on edited voxels too, leaving the surviving edit signal as whatever the optimiser could not capture.

\paragraph{Extrapolation between subsample means.}
Recall that $v_{\Delta}$~\eqref{eq:vdelta} is estimated via the shared noise $\varepsilon$ in the coupling~\eqref{eq:flowedit-coupling}. Each independent draw $\varepsilon^{(s)}$ produces a velocity difference $v_{\Delta}^{(s)}$, which we decompose as $v_{\Delta}^{(s)}=\mu+\eta^{(s)}$ where $\mu$ is the conditional expectation (the true edit signal) and $\eta^{(s)}$ is a zero-mean noise term. The $S$-sample mean $\hat\mu_{S}\!=\!\tfrac{1}{S}\sum_{s=1}^{S}v_{\Delta}^{(s)}$ and any partial mean $\hat\mu_{L}\!=\!\tfrac{1}{L}\sum_{s=1}^{L}v_{\Delta}^{(s)}$ with $1{\le}L{<}S$ are both unbiased for $\mu$ but with variances $\sigma_{\eta}^{2}/S$ and $\sigma_{\eta}^{2}/L$ respectively. Extrapolating away from the noisier estimate amplifies the direction in which the full-sample estimate is more confident:
\begin{equation}
\hat\mu_{S} \;+\; w\,(\hat\mu_{S}-\hat\mu_{L}) \;=\; (1+w)\,\hat\mu_{S} \;-\; w\,\hat\mu_{L},\qquad w\in\mathbb{R}_{\ge 0}.
\label{eq:pmg-means}
\end{equation}
\textbf{The gated behaviour emerges automatically.} On non-edited voxels RASI has driven $\mu$ toward zero, so $\hat\mu_{S}$ and $\hat\mu_{L}$ both hover near zero and their gap has no consistent direction; the extrapolation is negligible and identity is not perturbed. On edited voxels $\mu$ is large, $\hat\mu_{S}$ tracks it more faithfully than $\hat\mu_{L}$, and~\eqref{eq:pmg-means} lengthens exactly the edit-aligned component. The full Stage-1 update at active step $t_{k}$ becomes
\begin{equation}
z_{t_{k+1}}^{\mathrm{edit}} \;=\; z_{t_{k}}^{\mathrm{edit}} \;+\; \Delta t\,\big[\,\hat\mu_{S} \;+\; w\,(\hat\mu_{S}-\hat\mu_{L})\,\big],
\label{eq:final-update}
\end{equation}
where every $v_{\Delta}^{(s)}$ already carries the injected $\phi_{t}$ from RASI.

\subsection{Twin-Agreement Residual Injection for Sparse-Stage Preservation}
\label{sec:tar}

Stage~1 \textcolor{black}{determines} the edited coordinates $\mathcal{C}_{\mathrm{tgt}}$ but does not specify the Stage-2 geometry and Stage-3 material SLATs. \textcolor{black}{Running the sparse sampler under $c_{\mathrm{tgt}}$ can also alter non-edited tokens, as changes in the DINOv3 image-conditioning features propagate through cross-attention.} We resolve this by letting the sampler itself identify which tokens are edit-sensitive: a twin forward under $c_{\mathrm{src}}$ on the same scaffold and noise reveals, per token, whether the output changes; tokens that agree across the twin are pulled back toward the source encoding, while disagreeing tokens are left untouched.

\paragraph{Why FlowEdit coupling is unavailable.}
The coupling~\eqref{eq:flowedit-coupling} requires both ODE branches to share the same token space, yet $\mathcal{C}_{\mathrm{tgt}}$ and $\mathcal{C}_{\mathrm{src}}$ generally differ in voxel support after an add or remove edit, making direct coupling impossible. Even when supports overlap, \emph{coordinate coincidence does not imply content preservation}, which is why the geometry-based merge of Nano3D~\citep{nano3d} fails.

\paragraph{Twin-forward disagreement map.}
\textcolor{black}{\textbf{The key insight} is that tokens in non-edited regions should remain relatively stable when the condition changes from the source image to the target image, whereas edit-sensitive tokens should exhibit larger differences.} This lets the sampler itself draw the preserve/edit boundary without any external mask.

Let $\mathcal{S}_{\theta}^{\mathrm{stage}}$ denote the Stage-2 or Stage-3 sparse-DiT sampler. On the target scaffold $\mathcal{C}_{\mathrm{tgt}}$ we run twin forwards with identical seeded noise $\varepsilon$, schedule, and CFG weight---only the image embedding differs:
\begin{equation}
z_{\mathrm{tgt}} = \mathcal{S}_{\theta}^{\mathrm{stage}}\big(\mathcal{C}_{\mathrm{tgt}};\,\varepsilon,\,c_{\mathrm{tgt}}\big),\qquad z_{\mathrm{src}}^{\mathrm{twin}} = \mathcal{S}_{\theta}^{\mathrm{stage}}\big(\mathcal{C}_{\mathrm{tgt}};\,\varepsilon,\,c_{\mathrm{src}}\big).
\label{eq:tar-twin}
\end{equation}
Per-token disagreement $d_{i}\!=\!\|z_{\mathrm{tgt}}[i]-z_{\mathrm{src}}^{\mathrm{twin}}[i]\|_{2}$ is mapped to a preserve-confidence via robust quantile clipping:
$p_{\mathrm{flow}}[i]=1-\mathrm{clip}\big((d_{i}-q_{\alpha}(d))/(q_{\beta}(d)-q_{\alpha}(d)),\,0,\,1\big)$ with $(\alpha,\beta){=}(0.05,0.95)$.
Tokens producing nearly identical features under both conditions receive $p_{\mathrm{flow}}\!\approx\!1$ and are safe to preserve.

\paragraph{Residual injection.}
Let $z_{\mathrm{src}}^{\mathrm{enc}}$ be the SC-VAE encoding of the source asset on $\mathcal{C}_{\mathrm{src}}$ and $\mathcal{I}=\mathcal{C}_{\mathrm{tgt}}\cap\mathcal{C}_{\mathrm{src}}$ the integer-voxel intersection. On $\mathcal{I}$ we form a norm-clipped residual $r_{i}=\mathrm{clip}_{\tau}(z_{\mathrm{src}}^{\mathrm{enc}}[i]-z_{\mathrm{tgt}}[i])$ and blend:
\begin{equation}
z[i] \;=\; z_{\mathrm{tgt}}[i] \;+\; \lambda\,p_{\mathrm{flow}}[i]\,\mathbf{1}[p_{\mathrm{flow}}[i]\!\ge\!\vartheta]\,r_{i}\quad (i\in\mathcal{I}),\qquad z[i]=z_{\mathrm{tgt}}[i]\quad (i\in\mathcal{C}_{\mathrm{tgt}}\setminus\mathcal{I}),
\label{eq:tar-blend}
\end{equation}
with scale $\lambda\!\in\!(0,1]$ and threshold $\vartheta\!\in\![0,1)$. Edit-only tokens ($i\!\in\!\mathcal{C}_{\mathrm{tgt}}\setminus\mathcal{I}$) stay at the target branch, protecting fresh geometry or material; on intersection tokens the indicator zeroes the update wherever the twin disagrees, while agreeing tokens are softly retracted toward the source encoding with strength governed by $p_{\mathrm{flow}}$.

We apply~\eqref{eq:tar-twin}--\eqref{eq:tar-blend} once per SLAT (geometry and material). Since the main target forward is bit-identical to the $c_{\mathrm{tgt}}$ twin, \textbf{TAR adds exactly one extra sparse-DiT forward per stage}. The full pipeline (RASI + PMG on Stage~1, TAR on Stages~2--3) is given in Algorithm~\ref{alg:full} (Appendix).

\section{Experiments}
\label{sec:exp}

\textcolor{black}{We evaluate VS3D on an Objaverse-derived benchmark comprising 50 assets and 50 edit cases. Benchmark construction, metrics, and implementation details are provided in Appendix~\ref{app:setup}. VS3D uses a fixed set of hyperparameters across all edits in this benchmark.}

\subsection{Baselines}
\label{sec:baselines}

\paragraph{Baselines.}
We compare against six methods grouped by whether they require a user-supplied 3D mask.
\emph{Mask-based:}
\textbf{TRELLIS~2.0}~\citep{trellis2} (mask-conditioned native repaint),
\textbf{VoxHammer}~\citep{voxhammer} (inversion + KV-replacement),
\textbf{Instant3dit}~\citep{instant3dit} (feed-forward multi-view inpainting, 2D-lift), and
\textbf{VecSet-Edit}~\citep{vecsetedit} (training-free editing on TripoSG).
All mask-based methods receive the same human-annotated mask.
\emph{Mask-free:}
\textbf{Nano3D}~\citep{nano3d} (FlowEdit + connectivity-based Voxel/SLat-Merge) and
\textbf{Edit360}~\citep{edit360} (video-diffusion 2D-edit propagation around a 360$^{\circ}$ trajectory).
For fair comparison, we re-implement VoxHammer and Nano3D, both originally built on TRELLIS~1.0, on the same TRELLIS~2.0 backbone.

\subsection{Main comparison}
\label{sec:main}

\textbf{Quantitative comparison.}
Table~\ref{tab:quant} summarises the results on our benchmark.
\textcolor{black}{VS3D achieves the highest PSNR, SSIM, DINO-I, and CLIP-T and the lowest LPIPS among the compared methods. Its CD is second only to that of the mask-based VoxHammer.}

{\color{black}
To assess generalization beyond our benchmark, we additionally evaluate VS3D on Edit3D-Bench. We rerun VoxHammer, Nano3D, and VS3D under the same evaluation protocol. VS3D ranks first on five of the six metrics; detailed results are reported in Appendix~\ref{app:edit3dbench}.}

{\color{black}In a user study with 21 participants on challenging editing cases, VS3D receives the highest preference share in comparisons with Nano3D and VoxHammer for edit fidelity, identity preservation, and overall quality; details are provided in Appendix~\ref{app:userstudy}.}

\begin{figure}[t]
\centering
\includegraphics[width=\textwidth]{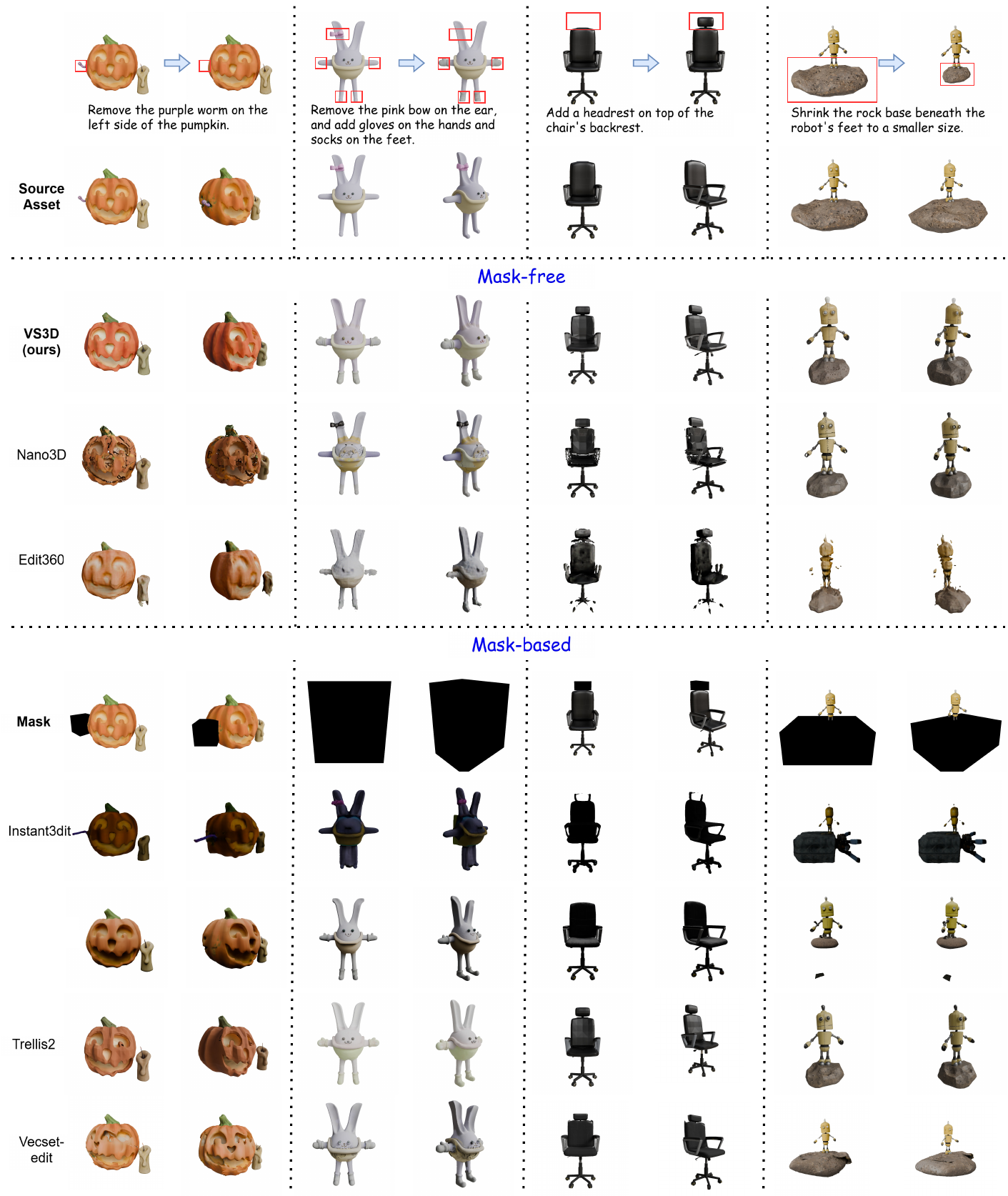}
\caption{\textbf{Qualitative comparison.} \textcolor{black}{Among the mask-free methods compared in Fig.~\ref{fig:qual}, ours is the only approach that jointly preserves the non-edited region and produces high-quality edits in the illustrated cases.} Red boxes highlight the edited regions for easier visual inspection and are not part of the model input. In the Mask row, black regions indicate the user-provided editing mask required by mask-based methods.}
\label{fig:qual}
\end{figure}

\textbf{Qualitative comparison.}
As shown in Fig.~\ref{fig:qual}, our method produces clean edits while faithfully preserving the geometry and texture of non-edited regions.
For elementary operations such as \emph{add} and \emph{remove}, our framework already achieves high-fidelity results with precise identity preservation.
\textcolor{black}{More importantly, it handles compound edits that simultaneously add and remove content, where the compared baselines struggle to achieve both the requested changes and source preservation.}
For \emph{deformation}-style edits that resemble replacement, our method also delivers competitive results, maintaining structural coherence throughout the transformation.
Additional large-scale qualitative results are provided in Appendix~\ref{app:gallery}.

\begin{table}[t]
\centering
\small
\vspace{-4pt}
\caption{\textbf{Quantitative comparison} on our Objaverse-derived editing benchmark. All metrics are averaged over the same set of edits. The \emph{Mask-free} column marks methods that do \emph{not} require a user-supplied 3D mask (\checkmark) versus those that do ($\times$). Best per column is in \textbf{bold}.}
\label{tab:quant}
\begin{tabular}{lccccccc}
\toprule
Method & Mask-free & PSNR$\uparrow$ & SSIM$\uparrow$ & LPIPS$\downarrow$ & DINO-I$\uparrow$ & CLIP-T$\uparrow$ & CD$\downarrow$ \\
\midrule
TRELLIS~2.0~\citep{trellis2}   & $\times$   & 17.30 & 0.793 & 0.288 & 0.720 & 0.278 & 0.089 \\
VoxHammer~\citep{voxhammer}    & $\times$   & 18.12 & 0.786 & 0.155 & 0.809 & 0.283 & \textbf{0.024} \\
Nano3D~\citep{nano3d}          & \checkmark & 19.04 & 0.801 & 0.242 & 0.715 & 0.285 & 0.096 \\
Instant3dit~\citep{instant3dit}& $\times$   & 14.96 & 0.684 & 0.229 & 0.609 & 0.267 & 0.053 \\
Edit360~\citep{edit360}        & \checkmark & 16.74 & 0.830 & 0.299 & 0.501 & 0.245 & 0.173 \\
VecSet-Edit~\citep{vecsetedit} & $\times$   & 20.40 & 0.829 & 0.199 & 0.763 & 0.212 & 0.051 \\
\midrule
\textbf{Ours}                  & \checkmark & \textbf{22.51} & \textbf{0.867} & \textbf{0.145} & \textbf{0.841} & \textbf{0.294} & 0.034 \\
\bottomrule
\end{tabular}
\end{table}

\subsection{Ablation studies}
\label{sec:ablation}

\begin{figure}[t]
\centering
\includegraphics[width=.90\linewidth]{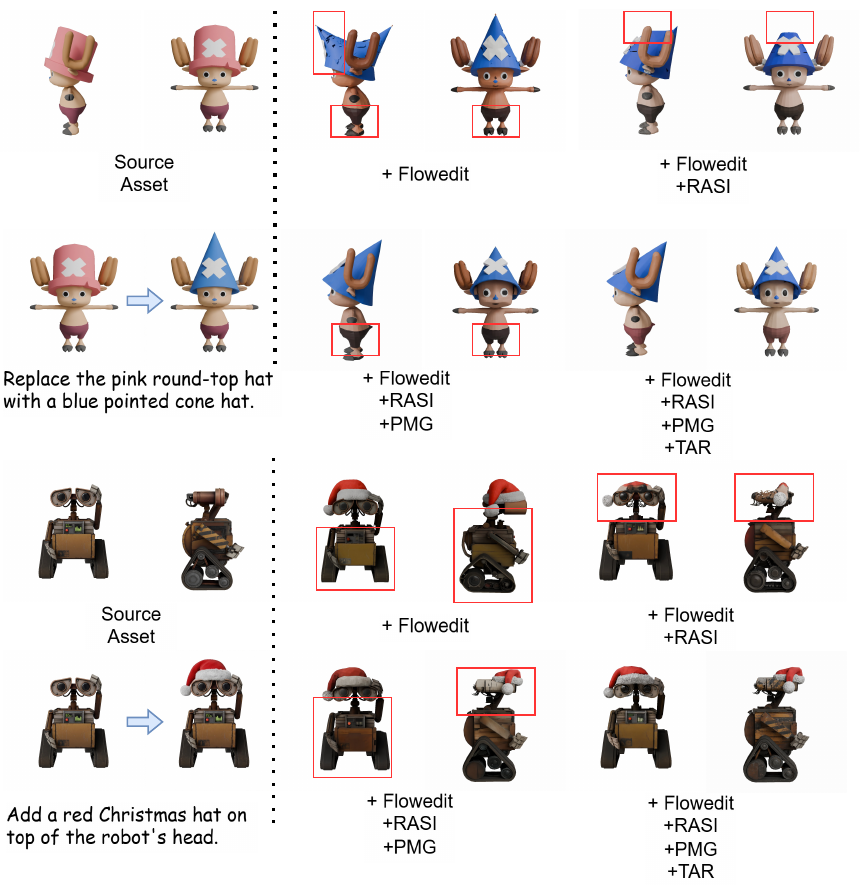}
\caption{\textbf{Ablation study.} Each module is progressively added and resolves a distinct failure mode.}
\label{fig:ablation}
\end{figure}

We ablate the proposed modules by progressively adding them on top of the base FlowEdit pipeline. Fig.~\ref{fig:ablation} shows two representative cases (rows~1--2 and rows~3--4). For each case, the first column displays the source asset (front and side views) together with the editing pair; subsequent columns incrementally activate each module: FlowEdit alone, +RASI, +RASI+PMG, and +RASI+PMG+TAR (full model).

\begin{figure}[t]
\centering
\small
\begin{tabular}{@{}c@{\hspace{.013\textwidth}}c@{\hspace{.013\textwidth}}c@{\hspace{.013\textwidth}}c@{}}
\textbf{Source} & \textbf{Target} & \textbf{Geometry map} & \textbf{Material map} \\
\includegraphics[width=.23\textwidth]{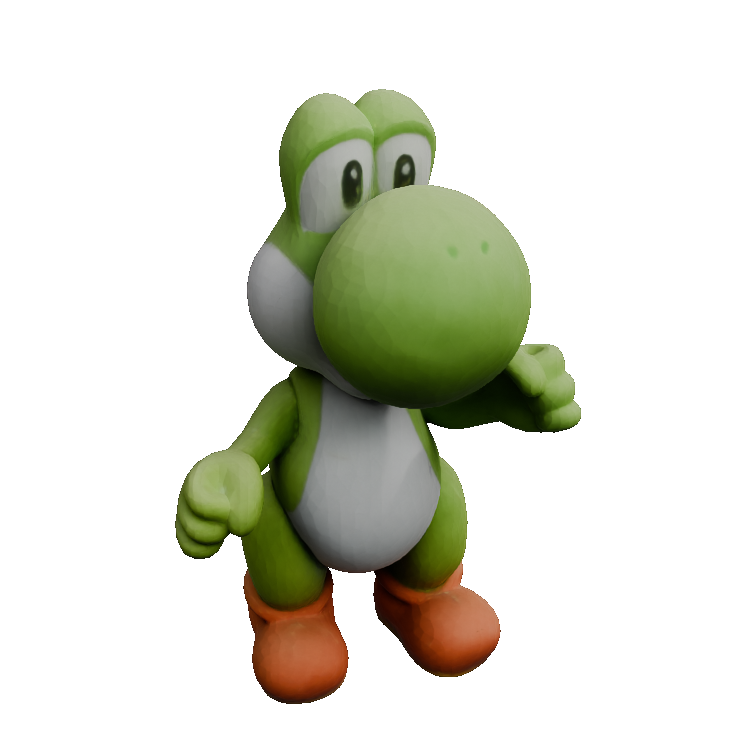} &
\includegraphics[width=.23\textwidth]{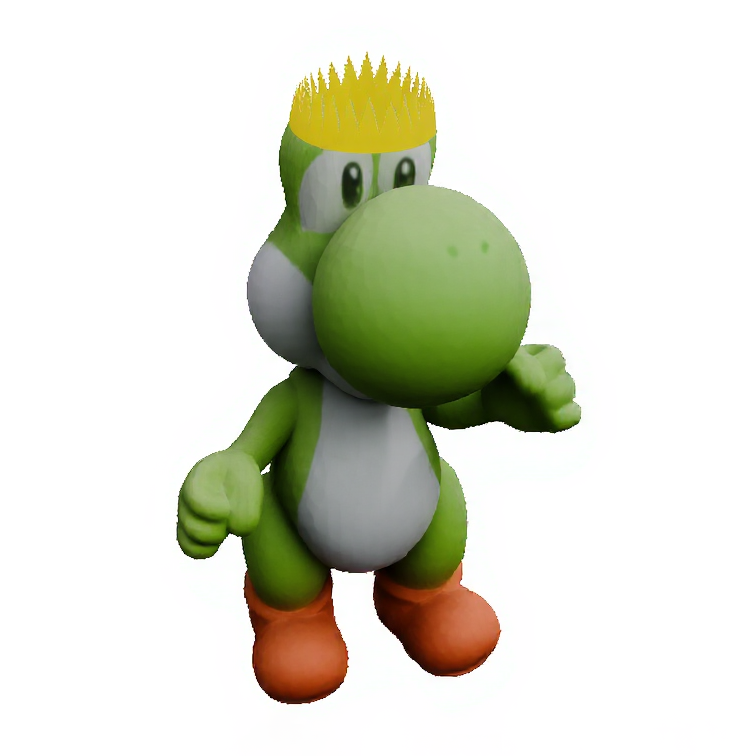} &
\includegraphics[width=.23\textwidth]{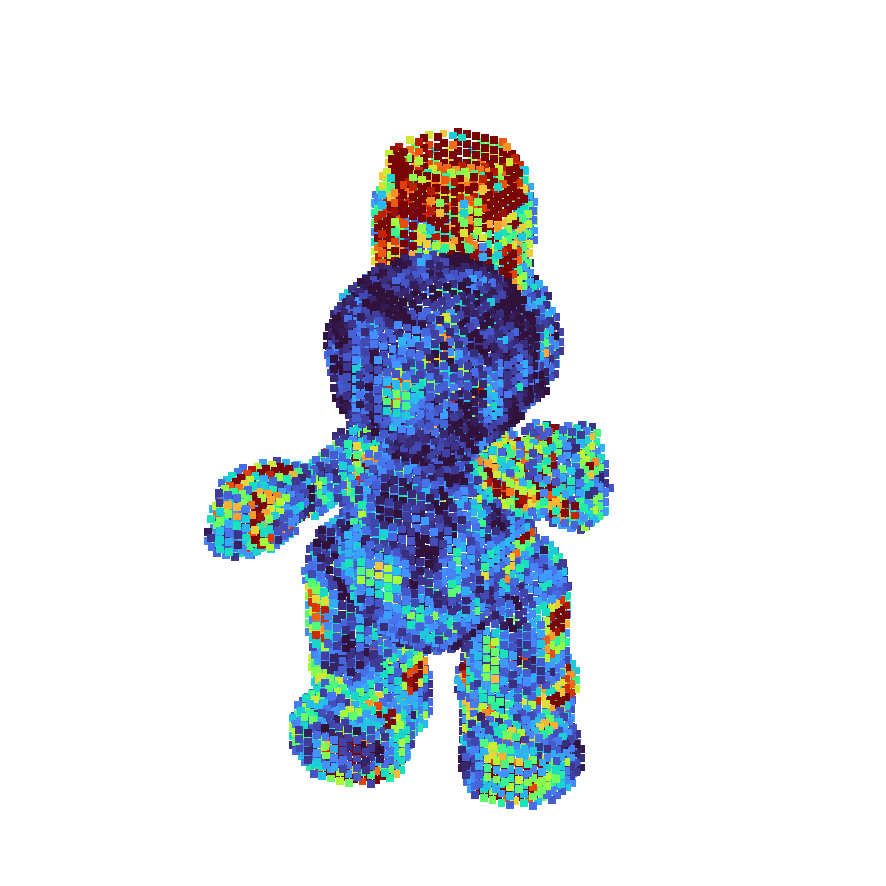} &
\includegraphics[width=.23\textwidth]{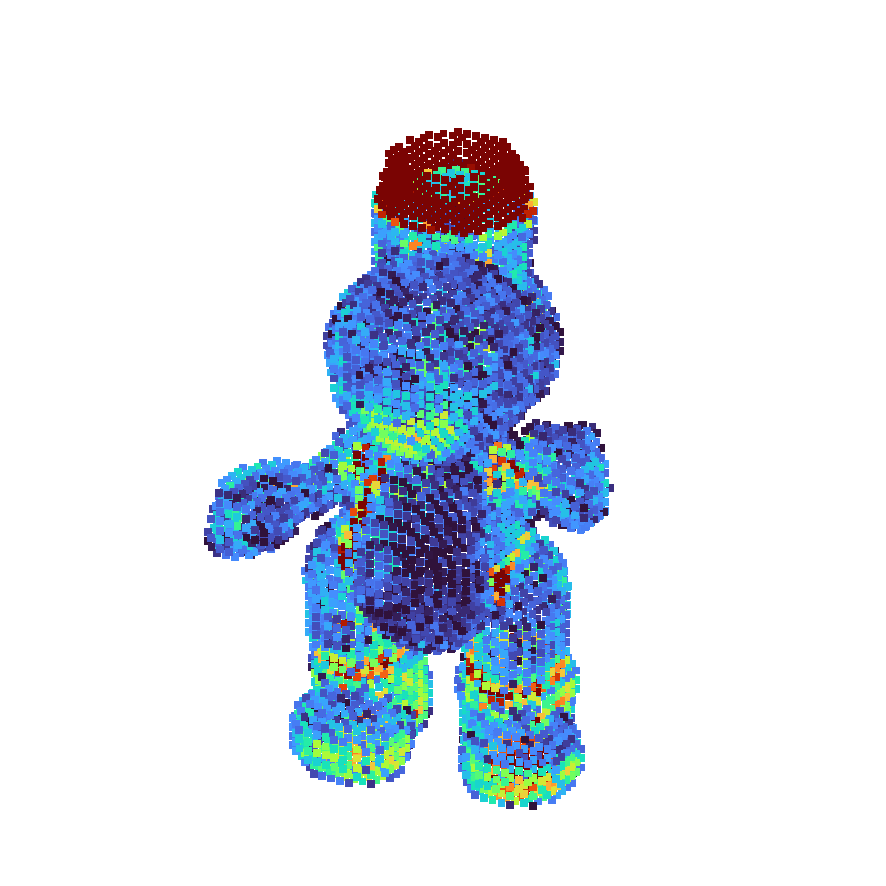} \\
\includegraphics[width=.23\textwidth]{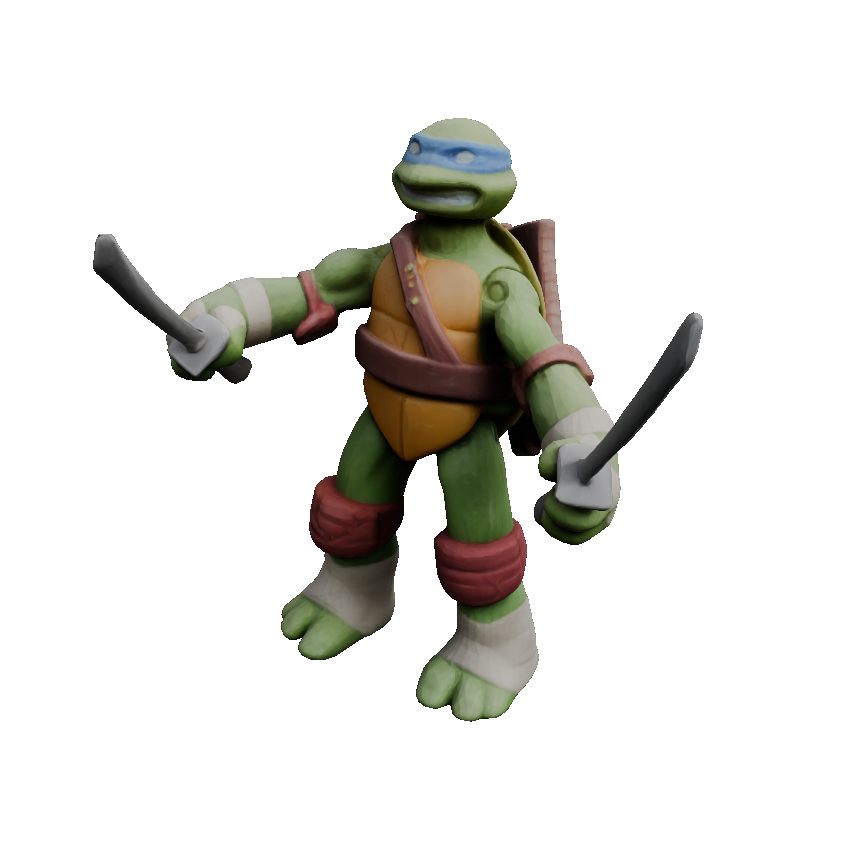} &
\includegraphics[width=.23\textwidth]{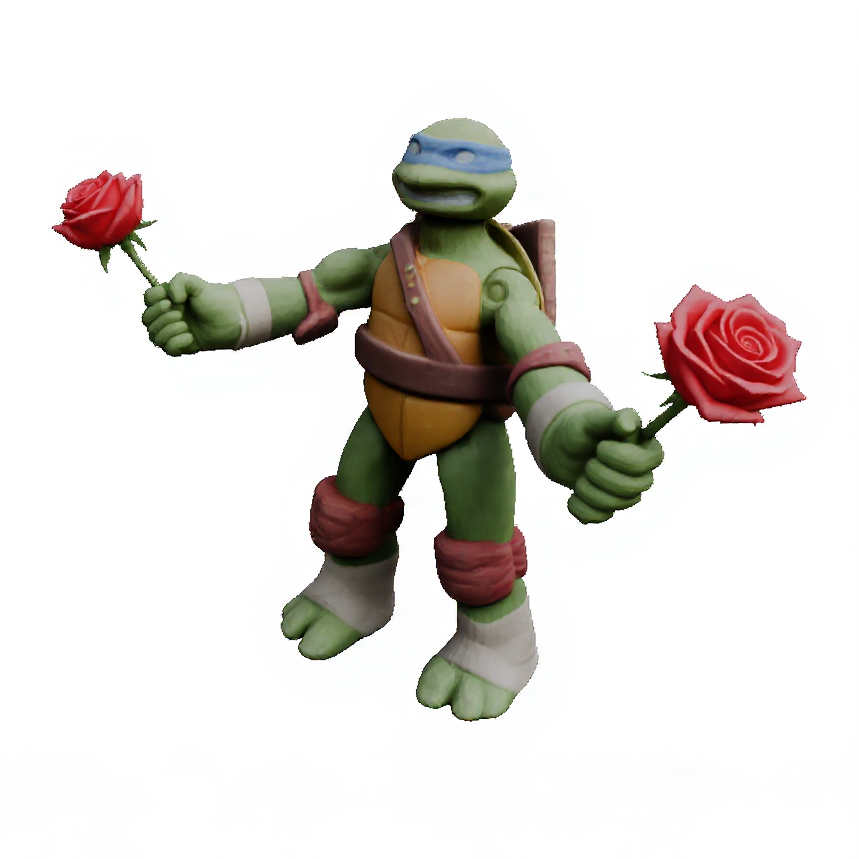} &
\includegraphics[width=.23\textwidth]{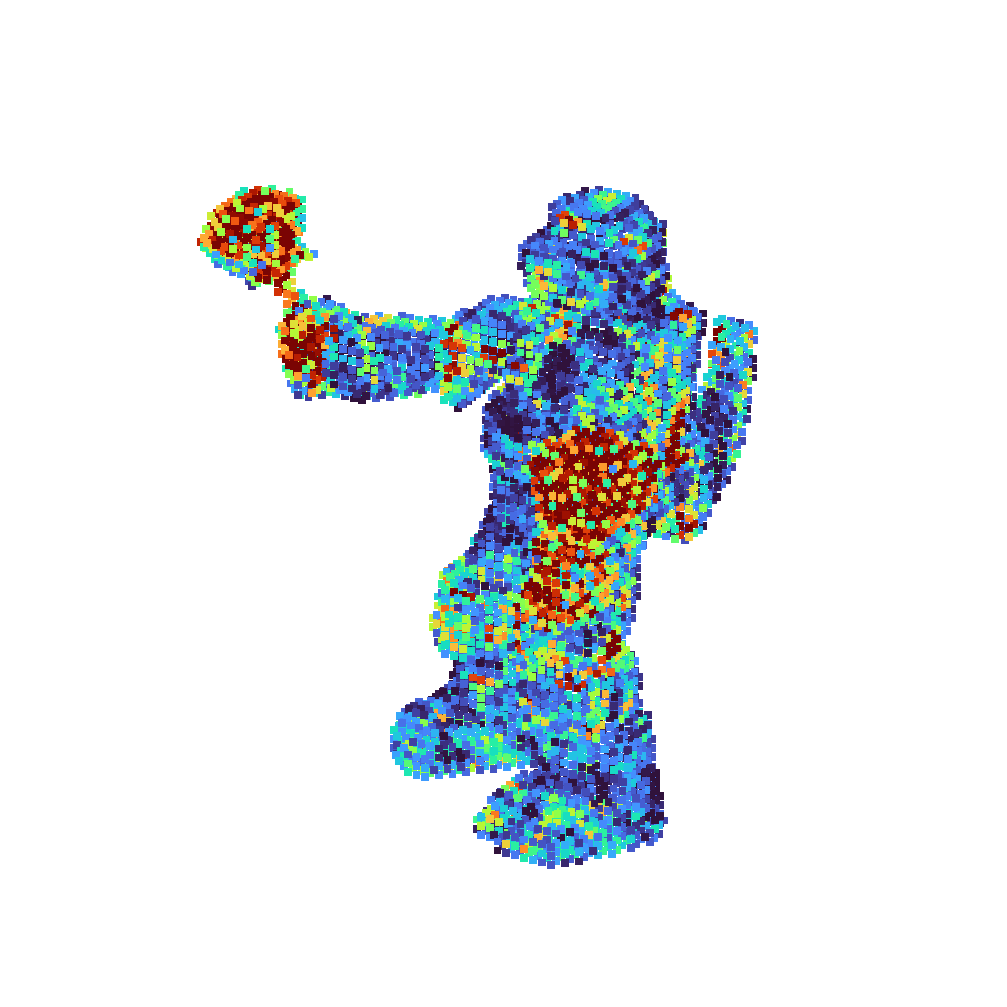} &
\includegraphics[width=.23\textwidth]{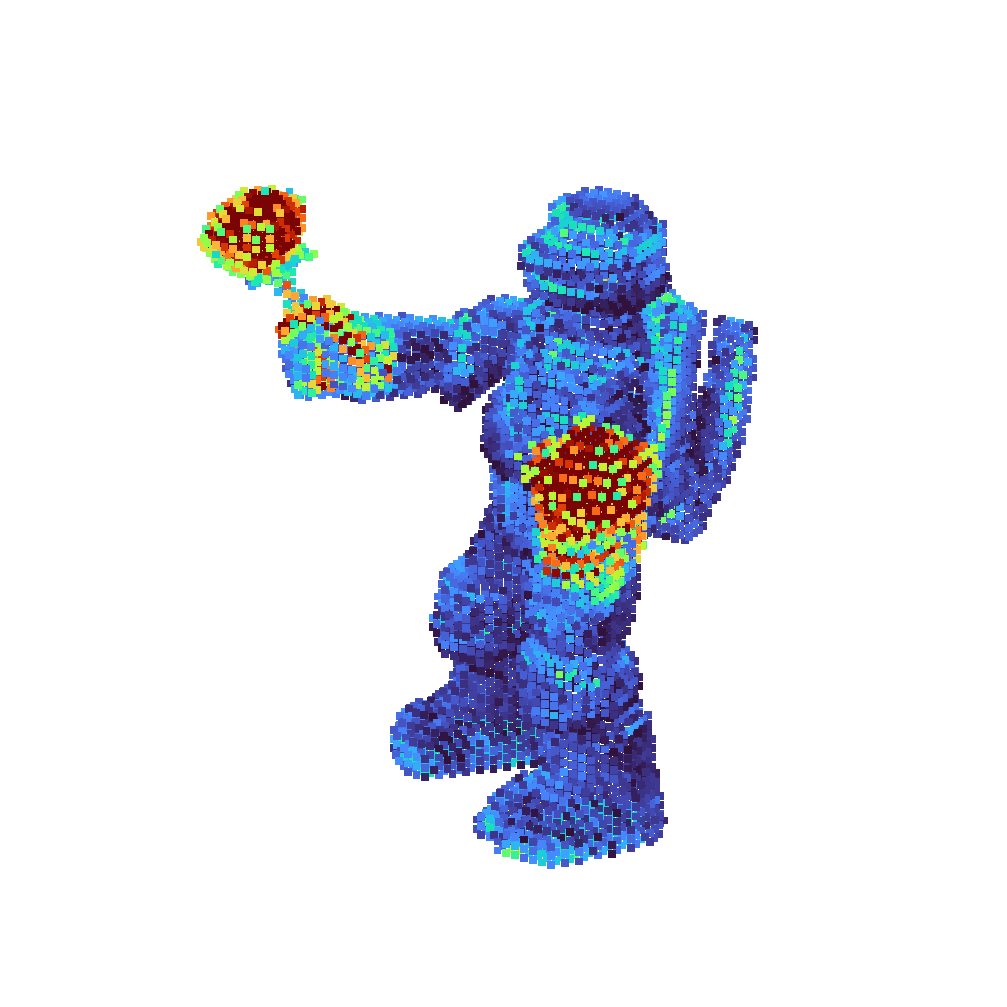} \\
\includegraphics[width=.23\textwidth]{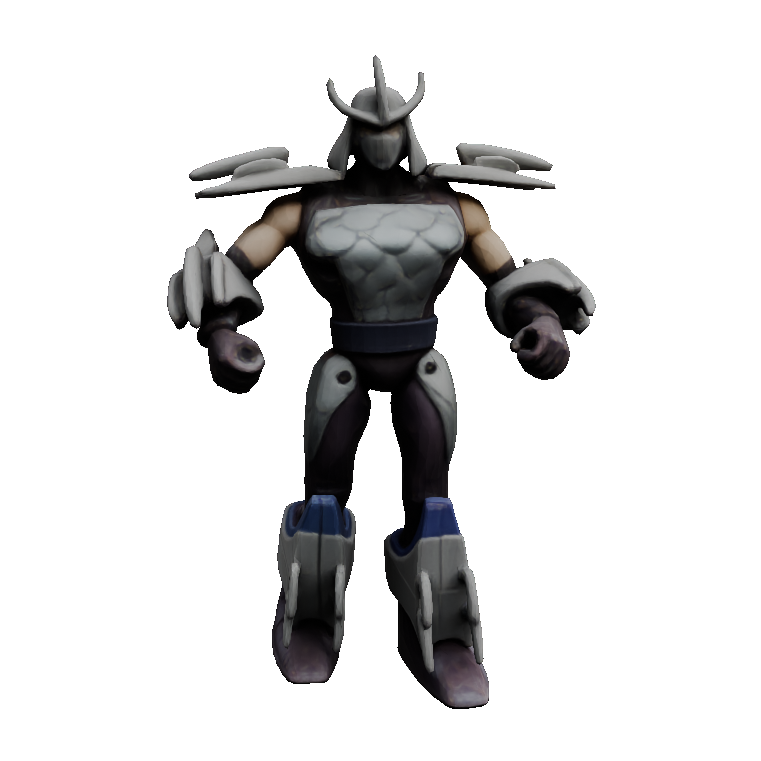} &
\includegraphics[width=.23\textwidth]{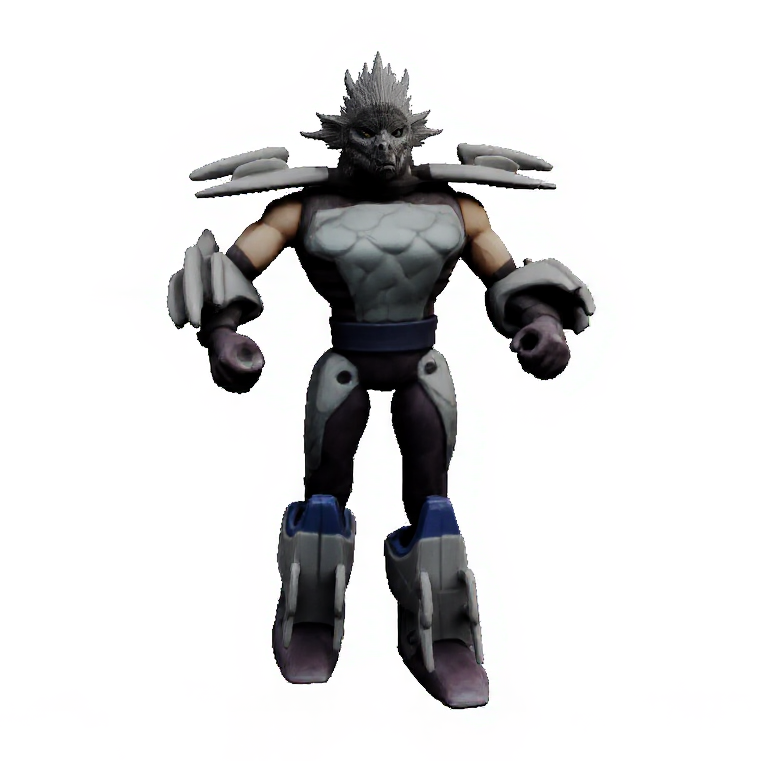} &
\includegraphics[width=.23\textwidth]{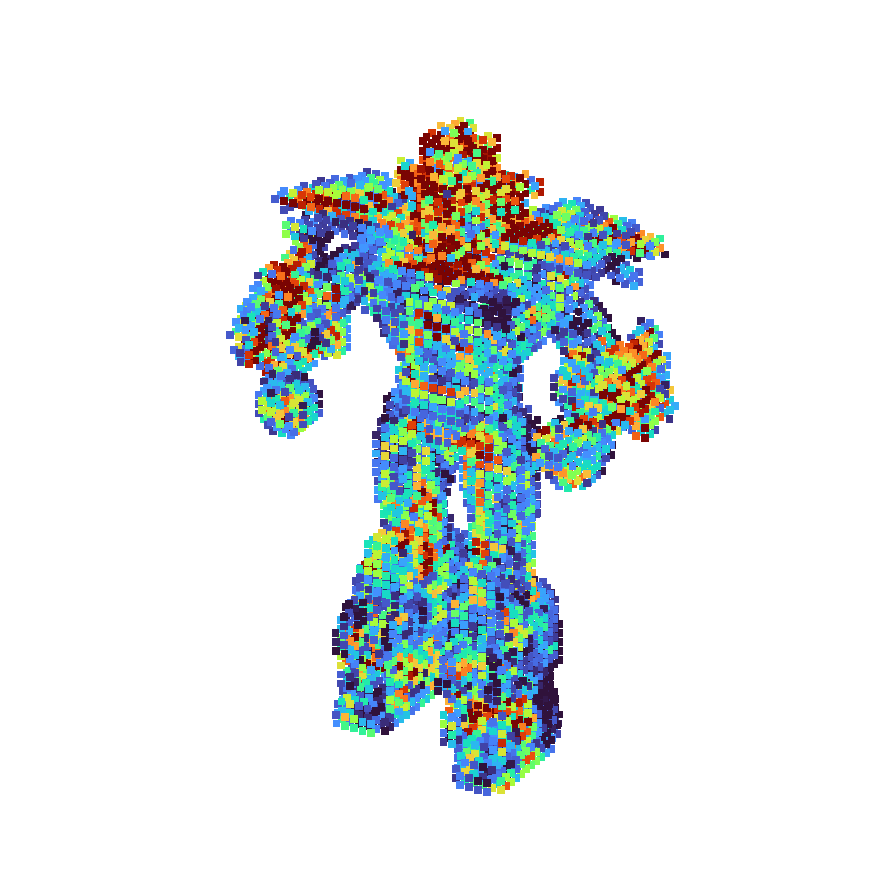} &
\includegraphics[width=.23\textwidth]{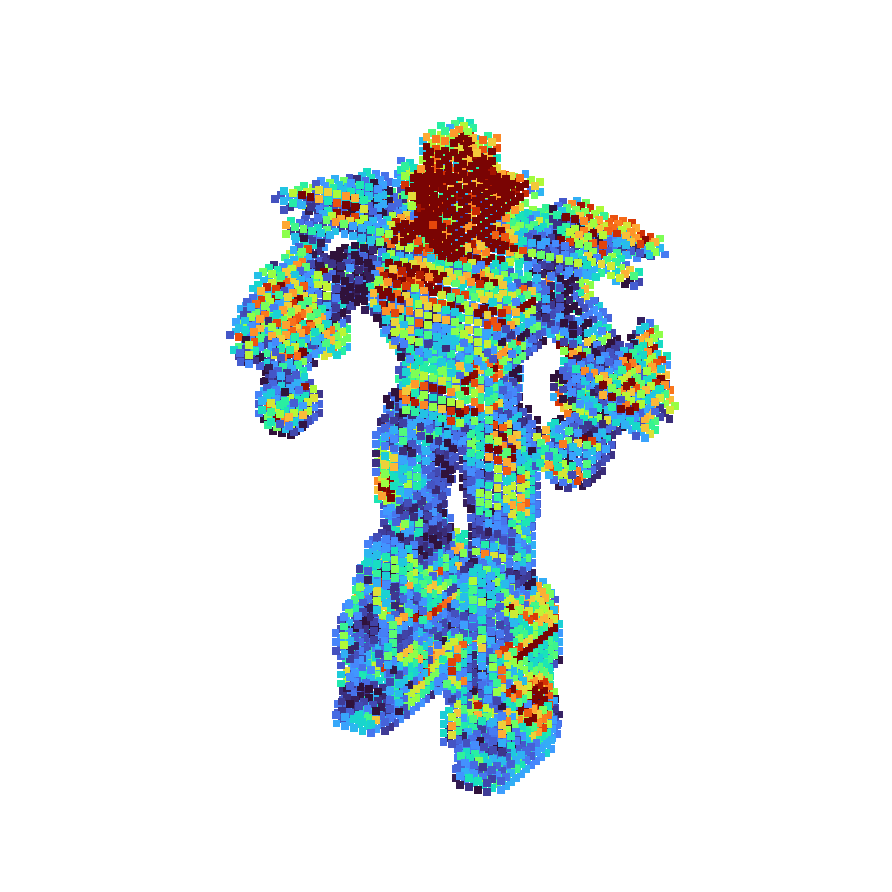}
\end{tabular}
\caption{\textcolor{black}{\textbf{TAR twin-branch disagreement maps on three Edit3D-Bench cases.} Rows show adding a crown, adding roses, and replacing a head with a dragon helmet. Columns show the source asset, the single-view target condition, and token-wise disagreement maps for geometry and material. Warmer colors indicate larger differences between source- and target-conditioned outputs; cooler colors indicate closer agreement. TAR uses this agreement to weight source-residual injection at overlapping source and target coordinates.}}
\label{fig:tar-main}
\end{figure}

FlowEdit alone suffers from severe voxel-occupancy drift on non-edited regions.
Adding RASI anchors the source trajectory and visibly corrects this drift---the moose's hat shape and position shift back toward the source, and the robot's torso largely recovers its original geometry---yet the edited region now lacks sufficient editing strength.
Introducing PMG amplifies the edit signal: the blue hat and the red Christmas hat both emerge clearly, while changes in non-edited areas remain well suppressed.
At this stage, however, fine-grained details (e.g.\ material and texture) of the non-edited region still deviate noticeably from the source, because Stages~2--3 are generated via forward sampling without explicit source-information injection.
Finally, TAR injects source-asset details into these later stages: the moose's trousers and body colour \textcolor{black}{more closely match the source}, and surface textures on the robot's torso are faithfully restored.
 {\color{black}Figure~\ref{fig:tar-main} visualizes the geometry and material twin-disagreement maps for three edits; additional cases are provided in Appendix~\ref{app:tar-vis}.}

{\color{black}To quantify the contribution of each module, we progressively add RASI, PMG, and TAR to the FlowEdit pipeline. Table~\ref{tab:ablation} reports the resulting performance: RASI produces the largest gains in preservation-related metrics, while PMG increases CLIP-T from 0.278 to 0.292. Adding TAR further improves PSNR, LPIPS, and CD over the RASI+PMG configuration, consistent with its role in the sparse geometry and material stages. The gains are not uniform across all metrics: RASI alone achieves the lowest LPIPS and CD and the highest DINO-I, whereas the full model achieves the highest PSNR and CLIP-T while matching the best SSIM. These results highlight the complementary roles of the modules and the trade-off between edit fidelity and source preservation.}

{\color{black}\begin{table}[H]
\color{black}
\centering
\small
\caption{\textbf{Quantitative ablation of the proposed modules.} Modules are added cumulatively to FlowEdit. Best values in each column are shown in bold.}
\label{tab:ablation}
\begin{tabular}{lcccccc}
\toprule
Configuration & PSNR$\uparrow$ & SSIM$\uparrow$ & LPIPS$\downarrow$ & DINO-I$\uparrow$ & CLIP-T$\uparrow$ & CD$\downarrow$ \\
\midrule
FlowEdit (baseline) & 20.18 & 0.804 & 0.166 & 0.813 & 0.225 & 0.063 \\
+ RASI & 22.01 & 0.856 & \textbf{0.142} & \textbf{0.848} & 0.278 & \textbf{0.032} \\
+ RASI + PMG & 22.43 & \textbf{0.867} & 0.146 & 0.845 & 0.292 & 0.036 \\
Full & \textbf{22.51} & \textbf{0.867} & 0.145 & 0.841 & \textbf{0.294} & 0.034 \\
\bottomrule
\end{tabular}
\end{table}}

\raggedbottom
\section{Discussion}
\label{sec:discussion}

VS3D achieves strong identity preservation in non-edit regions: RASI anchors the source trajectory to suppress occupancy drift, while PMG amplifies the edit signal only where needed.  {\color{black}In the geometry and material stages, TAR injects source-encoded residuals into tokens identified as suitable for preservation through twin-branch agreement. However, the inferred preserve/edit partition does not guarantee exact preservation: small changes in geometry or material may remain outside the intended edit region. VS3D is therefore intended for source-preserving local editing. Edits that replace nearly the entire asset fall outside this scope, as source anchoring may constrain the substantial changes required.}

\textcolor{black}{Furthermore, VS3D operates on frozen TRELLIS~2.0 models, and its editing quality depends on the backbone's ability to represent and generate the requested geometry and materials. Limitations of the backbone can therefore carry over to the edited results. Despite this dependence, VS3D demonstrates that high-quality shape and texture preservation can be achieved through ODE-space modifications without external 3D edit masks, inversion, or backbone fine-tuning. By combining source preservation with targeted editing in a training-free framework, VS3D offers a practical approach to local 3D asset editing. We hope this perspective motivates further research into source-preserving control of pretrained 3D generators.}

\section{Conclusion}
\label{sec:concl}

 {\color{black}We presented VS3D, a training-free, mask-free framework for high-fidelity local 3D asset editing that operates entirely within the velocity space of a frozen rectified-flow generator. By composing three lightweight ODE-level interventions---RASI for trajectory anchoring, PMG for selective edit amplification, and TAR for geometry and material preservation---VS3D achieves strong identity preservation and faithful editing without masks, inversion, or fine-tuning. Experiments on diverse object categories and edit types, together with evaluations on Edit3D-Bench and a user study, support its effectiveness.}

\clearpage
\flushbottom
\section*{References}

{\small
\renewcommand\bibsection{}

}


\appendix

\section{Technical appendices and supplementary material}

\subsection{Experimental setup}
\label{app:setup}

\paragraph{Benchmark.}
 \textcolor{black}{We evaluate on an Objaverse-derived benchmark containing 50 manually selected assets and 50 edit cases. The assets are stored in GLB format and cover characters, weapons, vehicles, creatures, furniture, and everyday objects. We select assets with complete geometry and topology, clear textures and materials, high render quality, and a well-defined local edit region, while maintaining category diversity. For each case, we manually write an edit instruction, render a source view that clearly exposes the target region, and use Nano Banana 2 to edit this view according to the instruction, producing the target condition image. Each case therefore consists of a source 3D asset, a source-view image, an edit instruction, and a single-view target condition image. The edit operations cover add, remove, replace, scale-up, shrink, and compound edits, with an approximate distribution of 20\% / 20\% / 20\% / 10\% / 10\% / 20\%. Since Objaverse is part of the multi-source corpus used by TRELLIS 2.0, overlap with some source assets is possible. However, all edit instructions and target condition images are newly created, so the evaluated asset-instruction-target tuples do not appear in pretraining. In addition, all TRELLIS-based baselines use the same source assets, edit instructions, target condition images, and TRELLIS 2.0 backbone, so any source-asset familiarity is shared across methods rather than favoring VS3D.}

\paragraph{Metrics.}
We report six metrics that jointly cover the three failure modes observed in 3D editing:
\textbf{(i)~PSNR$\uparrow$}, \textbf{(ii)~SSIM$\uparrow$}, and \textbf{(iii)~LPIPS$\downarrow$}~\citep{lpips}, measuring pixel-, structural-, and perceptual-level fidelity of the rendered views \textcolor{black}{against renderings of the source asset in the non-edited regions};
\textbf{(iv)~DINO-I$\uparrow$}, measuring multi-view semantic alignment to the target-view image via a DINOv2~\citep{dinov2} backbone;
\textbf{(v)~CLIP-T$\uparrow$}~\citep{clip}, measuring alignment between rendered views and the textual edit instruction;
\textbf{(vi)~CD$\downarrow$} (Chamfer Distance), measuring 3D geometric identity preservation against the source mesh on points sampled outside the human-annotated edit region.
PSNR/SSIM/LPIPS/DINO-I/CLIP-T are averaged over $16$ canonical views at $512^2$.

\paragraph{Implementation details.}
\emph{Stage-1 (RASI + PMG).} We use $T{=}25$ discretised timesteps, an active window governed by $n_{\max}{=}12$ and $n_{\min}{=}0$, and a noise-sample budget of $S{=}5$. PMG uses $(w,L)=(1.2,\,2)$. RASI uses $K{=}3$ inner steps with inner-loop learning rate $10^{-5}$ and early-stop threshold $10^{-5}$. The CFG weights are $\omega_{\mathrm{src}}{=}1.5$ and $\omega_{\mathrm{tgt}}{=}9.0$ over the CFG-active interval $t\!\in\![0.6,1.0]$.
\emph{Stage-2 / Stage-3 (TAR).} Both TAR stages run on the $1024$-token pipeline with its native sparse-DiT sampler defaults. The agreement field $p_{\mathrm{keep}}$ is computed with robust quantiles $(\alpha,\beta)=(0.05,0.95)$. \textcolor{black}{TAR in both the geometry and material stages uses} $\lambda{=}0.5$, norm-clip $\tau{=}10$, and agreement threshold $\vartheta{=}0.7$, with the residual anchored to the target branch (i.e.\ $r_{i}=\mathrm{clip}_{\tau}(z_{\mathrm{src}}^{\mathrm{enc}}[i]-z_{\mathrm{tgt}}[i])$). The src-encoded latents $z_{\mathrm{src}}^{\mathrm{enc}}$ are obtained by running the frozen SC-VAE encoders on the source GLB at $1024^{3}$; both Stage-2 and Stage-3 twins reuse the main-pipeline target forward for $z_{\mathrm{tgt}}$.

\subsection{Edit3D-Bench Evaluation}
\label{app:edit3dbench}

{\color{black}To assess performance beyond our Objaverse-derived benchmark, we evaluate VS3D on Edit3D-Bench and rerun VoxHammer and Nano3D under the same evaluation protocol. The results are independently reproduced rather than copied from the original papers.}

\begin{table}[H]
\color{black}
\centering
\small
\caption{Evaluation on Edit3D-Bench. All methods are evaluated under the same protocol.}
\label{tab:edit3dbench}
\begin{tabular}{lcccccc}
\toprule
Method & PSNR$\uparrow$ & SSIM$\uparrow$ & LPIPS$\downarrow$ & DINO-I$\uparrow$ & CLIP-T$\uparrow$ & CD$\downarrow$ \\
\midrule
VoxHammer & 13.8465 & 0.7195 & 0.4103 & \textbf{0.7014} & 0.2964 & 0.1969 \\
Nano3D & 15.2320 & 0.7544 & 0.4005 & 0.6959 & 0.2900 & 0.1931 \\
\textbf{VS3D (Ours)} & \textbf{15.3454} & \textbf{0.7621} & \textbf{0.3936} & 0.6999 & \textbf{0.2966} & \textbf{0.1834} \\
\bottomrule
\end{tabular}
\end{table}

{\color{black}VS3D ranks first on five of the six metrics, achieving the best PSNR, SSIM, LPIPS, CLIP-T, and CD. Its DINO-I is 0.0015 lower than VoxHammer.}

\subsection{Sensitivity to the 2D Editor}
\label{app:condition}

{\color{black}For each case, we render a single source view exposing the target region and edit it with the 2D editor to obtain the target condition image. This image specifies the intended edit, while all locality control, identity preservation, and geometry/material updates are performed within the 3D ODE. Thus, ``native 3D'' refers to where the edit is executed rather than how the target condition is authored. To evaluate sensitivity to the choice of 2D editor, we keep the source views, edit instructions, 3D pipeline, and hyper-parameters fixed, and only replace the editor used to produce the target condition.}

\begin{table}[H]
\color{black}
\centering
\small
\caption{Sensitivity to the choice of 2D editor used to produce the target condition. All other components are fixed.}
\label{tab:condition}
\begin{tabular}{lcccccc}
\toprule
2D editor & PSNR$\uparrow$ & SSIM$\uparrow$ & LPIPS$\downarrow$ & DINO-I$\uparrow$ & CLIP-T$\uparrow$ & CD$\downarrow$ \\
\midrule
\textbf{Nano Banana 2 (default)} & \textbf{22.51} & \textbf{0.867} & \textbf{0.145} & \textbf{0.841} & \textbf{0.294} & \textbf{0.034} \\
Nano Banana & 22.43 & 0.860 & 0.152 & 0.829 & 0.290 & 0.036 \\
Qwen-Image-Edit & 22.45 & 0.865 & 0.149 & 0.832 & 0.289 & 0.039 \\
\bottomrule
\end{tabular}
\end{table}

{\color{black}The results are close across the three editors: relative to the default, the largest deviations are 0.08\,dB in PSNR and 0.012 in DINO-I, while SSIM, LPIPS, CLIP-T, and CD vary by at most 0.007, 0.007, 0.005, and 0.005 respectively. This suggests that VS3D is not overly sensitive to the particular 2D editor used to author the condition.}

\subsection{Runtime comparison}
\label{app:runtime}

Table~\ref{tab:runtime} reports the wall-clock time for editing a single asset on a single NVIDIA RTX 4090 GPU.  {\color{black}VS3D completes the full three-stage pipeline in 57\,s without fine-tuning the backbone.}

\begin{table}[h]
\centering
\caption{Per-asset inference time (seconds) on a single NVIDIA RTX 4090 GPU.}
\label{tab:runtime}
\begin{tabular}{lcc}
\toprule
Method & Mask-free & Time (s) \\
\midrule
VoxHammer & \ding{55} & 394 \\
Instant3dit & \ding{55} & 46.2 \\
Nano3D & \ding{51} & 143 \\
Edit360 & \ding{51} & 495 \\
\textbf{VS3D (Ours)} & \ding{51} & \textbf{57} \\
\bottomrule
\end{tabular}
\end{table}

\subsection{User Study}
\label{app:userstudy}

{\small
{\color{black}We conducted a user study on challenging editing cases. We selected 20 assets, each paired with an edit instruction, including cases that require substantial structural modifications as well as compound edits that combine multiple operations. We recruited 21 participants, including 8 with backgrounds in 3D-related fields and 13 from other professional backgrounds. Participation was voluntary. Before taking part, participants were informed that the study involved viewing and comparing 3D-editing results; we anticipated no risks beyond viewing these outputs. Participants received no compensation.

Each participant evaluated all 20 cases. For each case, participants were shown the edit instruction, the source 2D rendering, the edited 2D target image, a 360-degree rendering video of the source 3D asset, and three anonymised result videos produced by VS3D, Nano3D, and VoxHammer. For each participant-case pair, the three results were randomly assigned the labels A, B, and C. Participants selected the best result separately according to three criteria: edit fidelity, identity preservation, and overall quality. Across 21 participants, 20 cases, and three criteria, the study produced 1{,}260 best-result choices.}

\begin{table}[H]
\color{black}
\centering
\small
\caption{User-study results. Each cell reports the number of best-result choices and the corresponding percentage out of 420 participant-case evaluations.}
\label{tab:userstudy}
\begin{tabular}{lccc}
\toprule
Criterion & \textbf{VS3D (Ours)} & Nano3D & VoxHammer \\
\midrule
Edit fidelity & \textbf{223 (53.1\%)} & 95 (22.6\%) & 102 (24.3\%) \\
Identity preservation & \textbf{218 (51.9\%)} & 73 (17.4\%) & 129 (30.7\%) \\
Overall quality & \textbf{223 (53.1\%)} & 82 (19.5\%) & 115 (27.4\%) \\
\bottomrule
\end{tabular}
\end{table}

{\color{black}VS3D is selected as the best result in a majority of participant-case evaluations under all three criteria. It leads the strongest baseline by 28.8, 21.2, and 25.7 percentage points for edit fidelity, identity preservation, and overall quality, respectively. These results provide complementary human-preference evidence on challenging edits.}
}

\paragraph{Broader impacts.}
\label{app:broader-impact}
\textcolor{black}{VS3D may reduce the manual effort required for local 3D asset editing in creative workflows. However, it could also be used to modify copyrighted assets without authorization or create misleading visual content. Users should respect asset licenses and obtain permission where required.}

\subsection{Full algorithm pseudocode}
\label{app:algorithm}

For completeness, Algorithm~\ref{alg:full} summarizes the complete VS3D inference pipeline. It follows the three-stage structure described in Sec.~3: RASI calibration and PMG editing are applied to dense occupancy latents in Stage~1, while TAR performs source-residual injection for geometry and material SLATs in Stages~2 and~3. The pseudocode makes explicit the stage-wise inputs, twin forwards, agreement-field computation, and final decoding process.

{\footnotesize
\begin{algorithm}[t]
\caption{VS3D: three-stage editing pipeline (RASI + PMG + TAR)}
\label{alg:full}
\begin{algorithmic}[1]
\Require source latent $x_{\mathrm{src}}$; source mesh/asset for SC-VAE encoding; conditions $c_{\mathrm{src}},c_{\mathrm{tgt}}$; active timesteps $\mathcal{T}_{\mathrm{act}}$; CFG weights $\omega_{\mathrm{src}},\omega_{\mathrm{tgt}}$; noise samples $S$; PMG $(w,L)$; RASI $(K,\eta_{0},\tau_{\mathrm{es}})$; TAR $(\lambda,\tau,\vartheta)$.
\Statex
\State \Comment{\textbf{Stage 1: Occupancy editing (RASI + PMG)}}
\Statex \Comment{Phase 1: RASI calibration}
\State $z^{\mathrm{edit}} \gets x_{\mathrm{src}}$
\For{$t\in\mathcal{T}_{\mathrm{act}}$ in descending order}
    \State initialise $\phi_{t}\gets\phi_{0}$; set $\eta_{t}$ by linear anneal
    \For{$K$ inner steps}
        \State form $z_{t}^{\mathrm{src}},z_{t}^{\mathrm{tgt}}$ via~\eqref{eq:flowedit-coupling}; evaluate $\mathcal{L}_{\mathrm{rec}}(\phi_{t})$ of~\eqref{eq:rasi-obj}
        \State $\phi_{t}\gets\mathrm{Adam}(\phi_{t},\nabla_{\phi_{t}}\mathcal{L}_{\mathrm{rec}},\eta_{t})$; \textbf{break} if $\mathcal{L}_{\mathrm{rec}}<\tau_{\mathrm{es}}$
    \EndFor
    \State advance $z^{\mathrm{edit}}$ by one Euler step on the source-reconstruction ODE; cache $\phi_{t}$
\EndFor
\Statex \Comment{Phase 2: PMG editing}
\State $z^{\mathrm{edit}} \gets x_{\mathrm{src}}$
\For{$t\in\mathcal{T}_{\mathrm{act}}$ in descending order}
    \For{$s=1,\ldots,S$}
        \State sample $\varepsilon^{(s)}\sim\mathcal{N}(0,\mathbf{I})$; form $z_{t}^{\mathrm{src}},z_{t}^{\mathrm{tgt}}$ via~\eqref{eq:flowedit-coupling}
        \State compute $\tilde v_{\mathrm{src}},\tilde v_{\mathrm{tgt}}$ with CFG using cached $\phi_{t}$; set $v_{\Delta}^{(s)}=\tilde v_{\mathrm{tgt}}-\tilde v_{\mathrm{src}}$
    \EndFor
    \State $\hat\mu_{S}\gets\tfrac{1}{S}\sum_{s=1}^{S}v_{\Delta}^{(s)}$;\; $\hat\mu_{L}\gets\tfrac{1}{L}\sum_{s=1}^{L}v_{\Delta}^{(s)}$
    \State $u \gets \hat\mu_{S} + w\,(\hat\mu_{S}-\hat\mu_{L})$ \Comment{PMG contrast}
    \State $z^{\mathrm{edit}}\gets z^{\mathrm{edit}} + \Delta t\,u$
\EndFor
\State decode $z^{\mathrm{edit}}$ to edited occupancy $\mathcal{C}_{\mathrm{tgt}}$
\Statex
\State \Comment{\textbf{Stage 2: Geometry SLAT (TAR)}}
\State $z_{\mathrm{tgt}}^{\mathrm{geo}} \gets \mathcal{S}_{\theta}^{\mathrm{geo}}(\mathcal{C}_{\mathrm{tgt}};\,\varepsilon,\,c_{\mathrm{tgt}})$ \Comment{native sparse-DiT forward}
\State $z_{\mathrm{src}}^{\mathrm{geo,twin}} \gets \mathcal{S}_{\theta}^{\mathrm{geo}}(\mathcal{C}_{\mathrm{tgt}};\,\varepsilon,\,c_{\mathrm{src}})$ \Comment{condition-swapped twin}
\State compute per-token agreement $p_{\mathrm{keep}}^{\mathrm{geo}}$ via robust quantile clipping on $\|z_{\mathrm{tgt}}^{\mathrm{geo}}[i]-z_{\mathrm{src}}^{\mathrm{geo,twin}}[i]\|_{2}$
\State $z_{\mathrm{src}}^{\mathrm{geo,enc}} \gets \mathrm{SC\text{-}VAE}_{\mathrm{geo}}(\text{source asset on } \mathcal{C}_{\mathrm{src}})$
\State blend via~\eqref{eq:tar-blend} with $(\lambda,\tau,\vartheta)$ on $\mathcal{I}=\mathcal{C}_{\mathrm{tgt}}\cap\mathcal{C}_{\mathrm{src}}$ to obtain $z^{\mathrm{geo}}$
\Statex
\State \Comment{\textbf{Stage 3: Material SLAT (TAR)}}
\State $z_{\mathrm{tgt}}^{\mathrm{mat}} \gets \mathcal{S}_{\theta}^{\mathrm{mat}}(\mathcal{C}_{\mathrm{tgt}},z^{\mathrm{geo}};\,\varepsilon',\,c_{\mathrm{tgt}})$ \Comment{geometry-conditioned}
\State $z_{\mathrm{src}}^{\mathrm{mat,twin}} \gets \mathcal{S}_{\theta}^{\mathrm{mat}}(\mathcal{C}_{\mathrm{tgt}},z^{\mathrm{geo}};\,\varepsilon',\,c_{\mathrm{src}})$
\State compute per-token agreement $p_{\mathrm{keep}}^{\mathrm{mat}}$; encode $z_{\mathrm{src}}^{\mathrm{mat,enc}}$
\State blend via~\eqref{eq:tar-blend} to obtain $z^{\mathrm{mat}}$
\Statex
\State \Return $(\mathcal{C}_{\mathrm{tgt}},\, z^{\mathrm{geo}},\, z^{\mathrm{mat}})$ \Comment{decode to mesh via Flexicubes + texture}
\end{algorithmic}
\end{algorithm}
}

\clearpage
\raggedbottom
\subsection{TAR agreement-field visualisation}
\label{app:tar-vis}

\textcolor{black}{We visualise token-wise differences between the source- and target-conditioned branches in the geometry and material stages. Warmer colours indicate larger differences, whereas cooler colours indicate closer agreement. TAR uses this agreement to weight source-residual injection at overlapping source and target coordinates.}

\begin{figure}[H]
\centering
\small
{\color{black}\makebox[0.2275\textwidth]{\textbf{Source Image}}\makebox[0.2275\textwidth]{\textbf{Target Image}}\makebox[0.2275\textwidth]{\textbf{Geometry Difference}}\makebox[0.2275\textwidth]{\textbf{Material Difference}}}
\par\vspace{6pt}
\begin{minipage}{\textwidth}
\centering
\includegraphics[width=.91\textwidth]{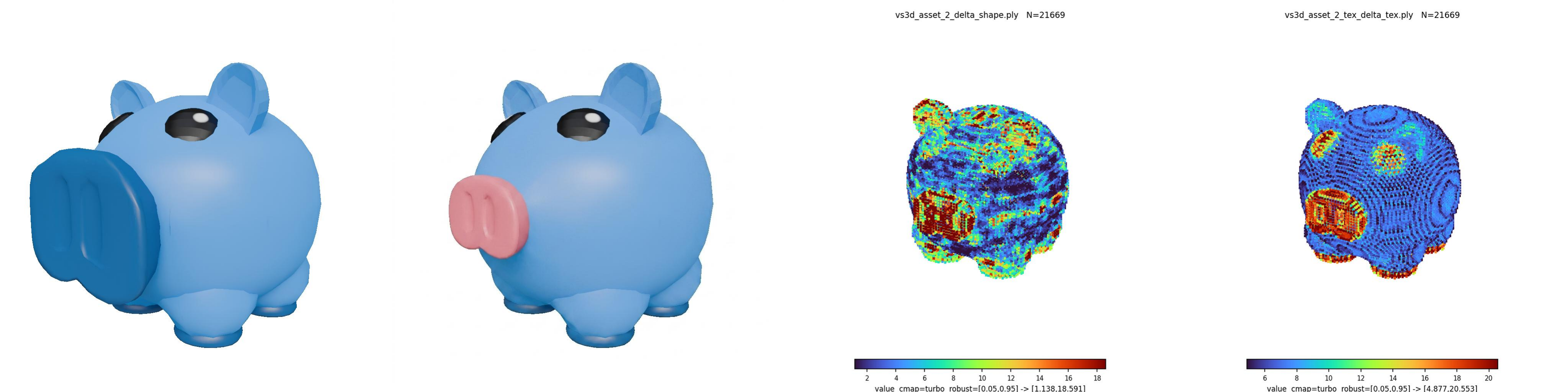}
\\[2pt]
{\small \emph{Replace the large dark blue pig snout with a smaller pink snout, keep everything else unchanged.}}
\end{minipage}
\par\vspace{4pt}
\begin{minipage}{\textwidth}
\centering
\includegraphics[width=.91\textwidth]{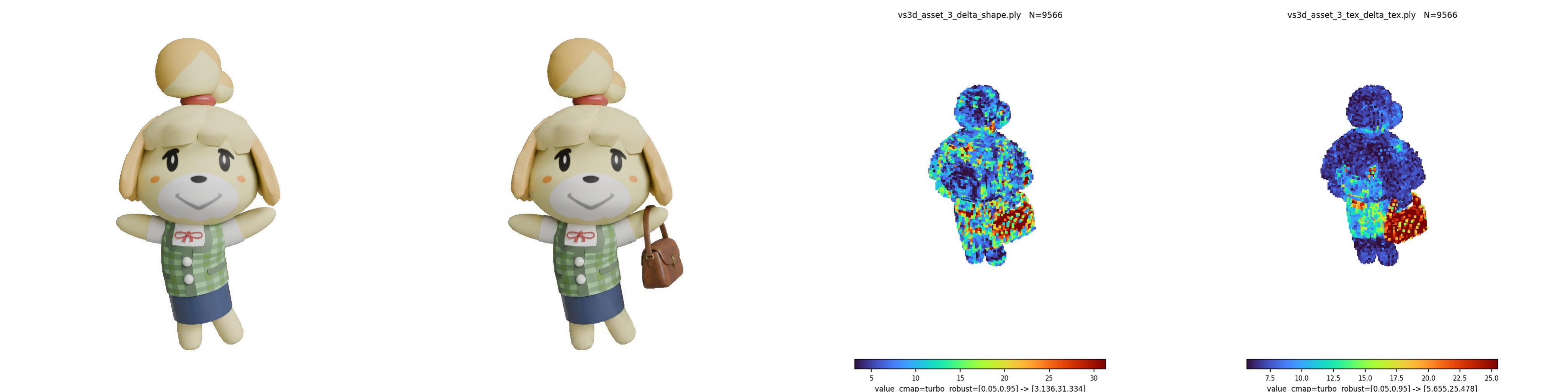}
\\[2pt]
{\small \emph{Add a small handbag to the character's hand, keep everything else unchanged.}}
\end{minipage}
\par\vspace{4pt}
\begin{minipage}{\textwidth}
\centering
\includegraphics[width=.91\textwidth]{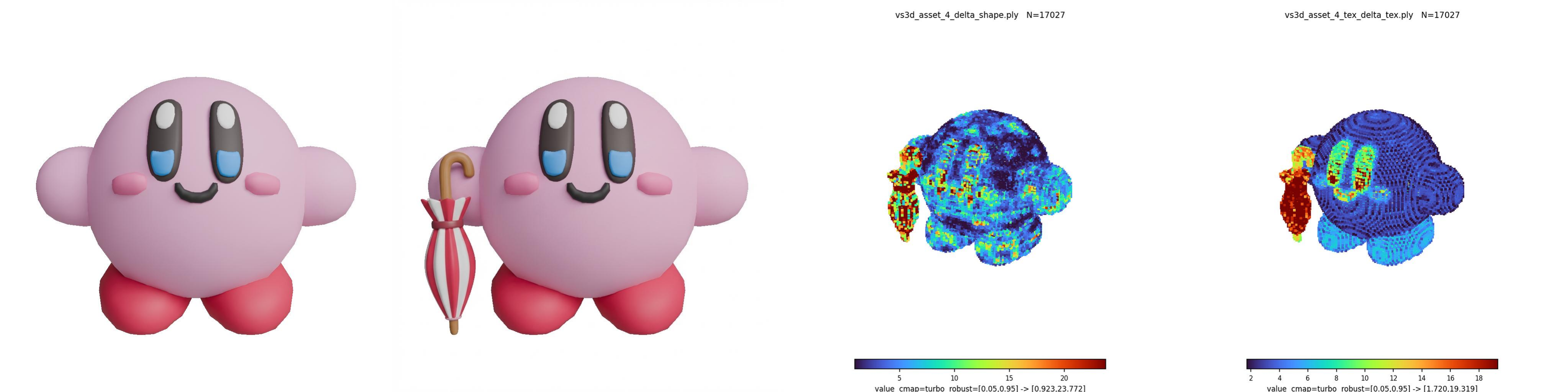}
\\[2pt]
{\small \emph{Add a closed red-and-white striped umbrella to the character's hand, keep everything else unchanged.}}
\end{minipage}
\par\vspace{4pt}
\begin{minipage}{\textwidth}
\centering
\includegraphics[width=.91\textwidth]{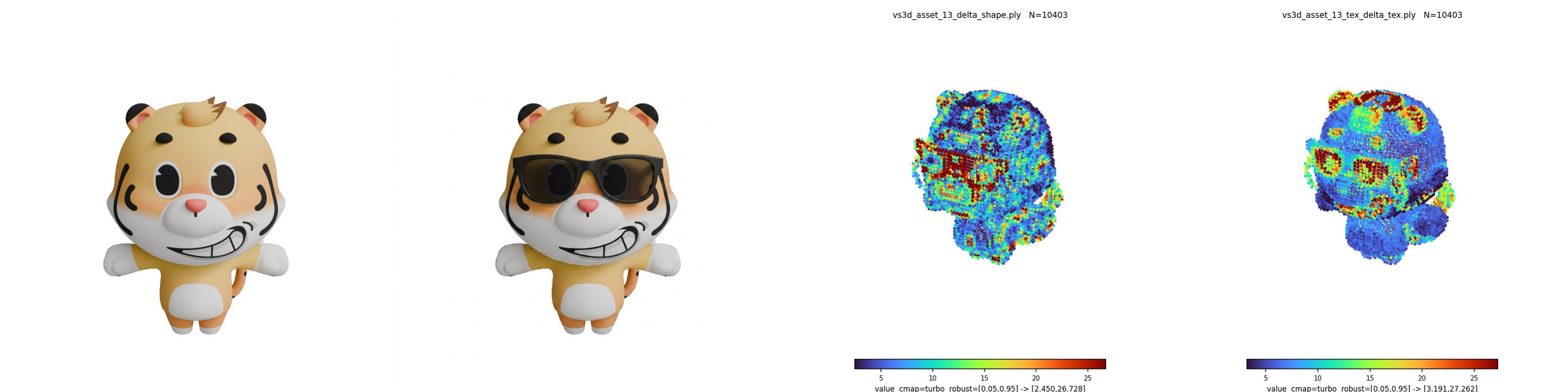}
\\[2pt]
{\small \emph{Add a pair of black sunglasses on the character's face, keep everything else unchanged.}}
\end{minipage}
\caption{\textcolor{black}{\textbf{TAR twin-branch disagreement maps (Part 1).} Columns show the source image, target image, geometry disagreement, and material disagreement. Warmer colours indicate larger differences between the source- and target-conditioned outputs; cooler colours indicate closer agreement. TAR uses this agreement to weight source-residual injection at overlapping source and target coordinates.}}
\label{fig:tar-vis-1a}
\end{figure}

\clearpage
\begin{figure}[H]
\centering
\small
{\color{black}\makebox[0.2275\textwidth]{\textbf{Source Image}}\makebox[0.2275\textwidth]{\textbf{Target Image}}\makebox[0.2275\textwidth]{\textbf{Geometry Difference}}\makebox[0.2275\textwidth]{\textbf{Material Difference}}}
\par\vspace{6pt}
\begin{minipage}{\textwidth}
\centering
\includegraphics[width=.91\textwidth]{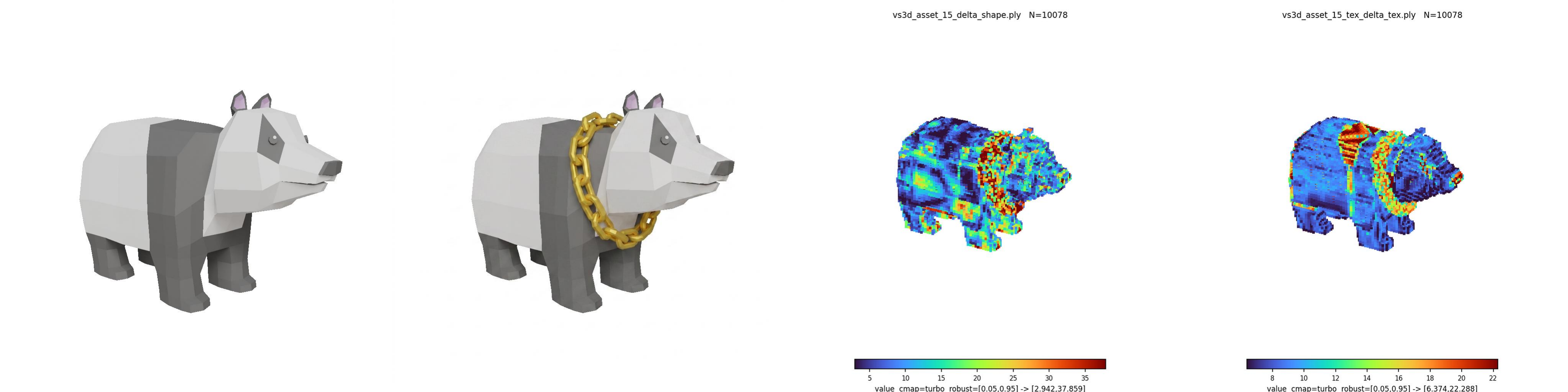}
\\[2pt]
{\small \emph{Add a big gold chain necklace around the character's neck, keep everything else unchanged.}}
\end{minipage}
\par\vspace{4pt}
\begin{minipage}{\textwidth}
\centering
\includegraphics[width=.91\textwidth]{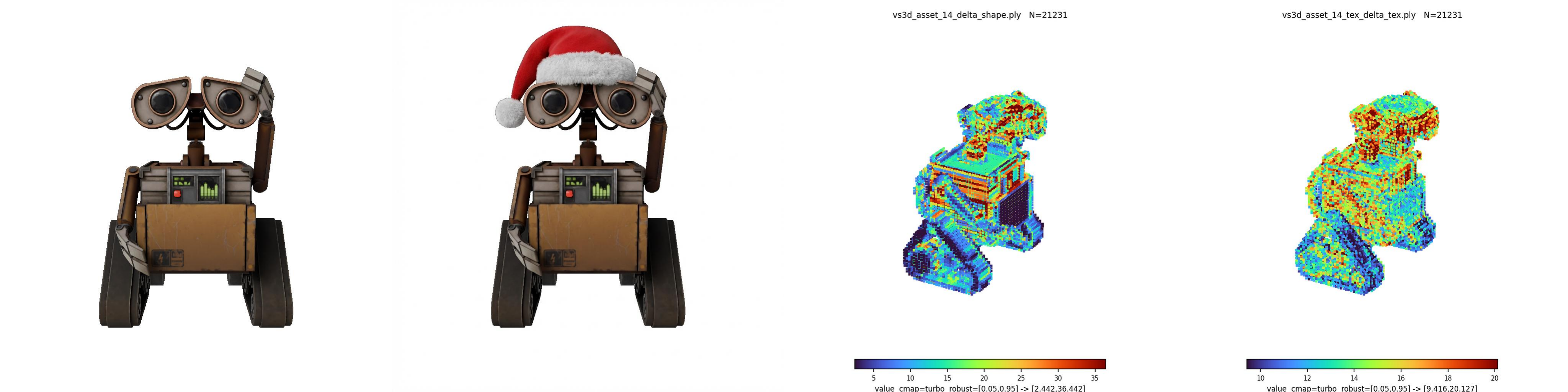}
\\[2pt]
{\small \emph{Add a red Christmas hat on top of the robot's head, keep everything else unchanged.}}
\end{minipage}
\par\vspace{4pt}
\begin{minipage}{\textwidth}
\centering
\includegraphics[width=.91\textwidth]{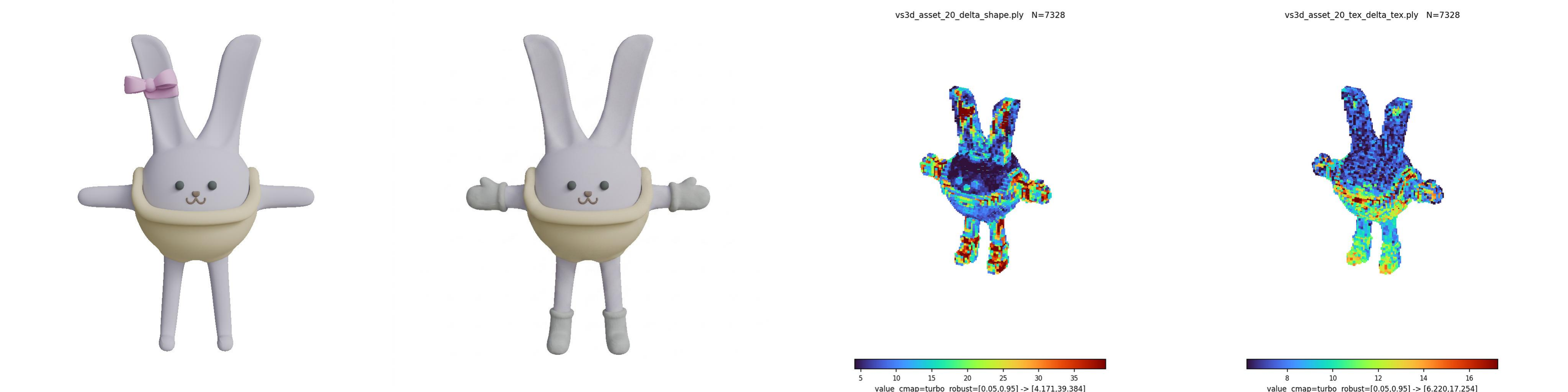}
\\[2pt]
{\small \emph{Remove the pink bow on the rabbit's ear, add gloves on its hands and socks on its feet, keep everything else unchanged.}}
\end{minipage}
\par\vspace{4pt}
\begin{minipage}{\textwidth}
\centering
\includegraphics[width=.91\textwidth]{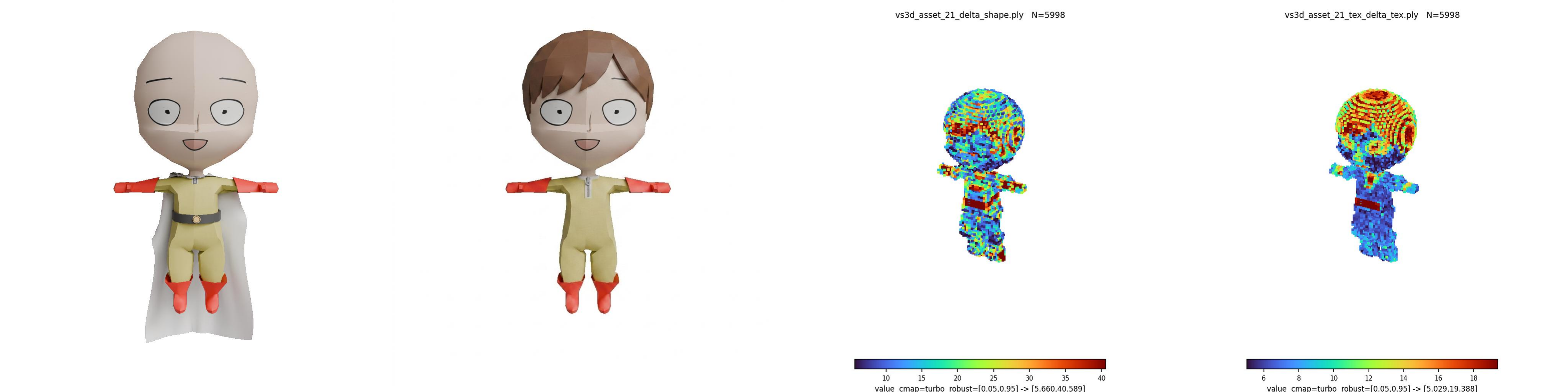}
\\[2pt]
{\small \emph{Remove the belt and the white cape from the character, and add side-swept bangs on the character's head, keep everything else unchanged.}}
\end{minipage}
\par\vspace{4pt}
\begin{minipage}{\textwidth}
\centering
\includegraphics[width=.91\textwidth]{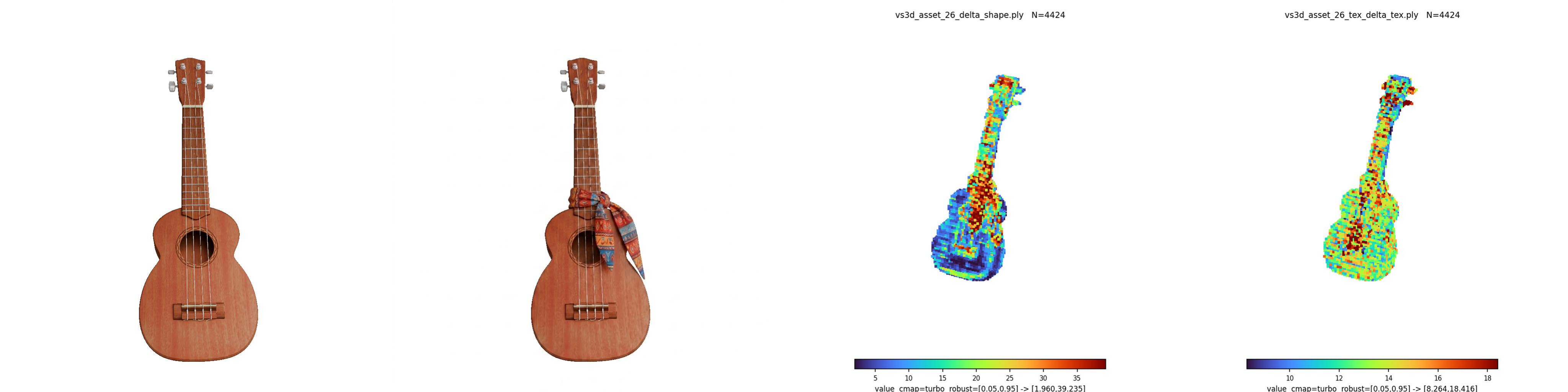}
\\[2pt]
{\small \emph{Add a scarf tied around the neck of the guitar, keep everything else unchanged.}}
\end{minipage}
\caption{\textcolor{black}{\textbf{TAR twin-branch disagreement maps (Part 2).} Columns show the source image, target image, geometry disagreement, and material disagreement. Warmer colours indicate larger differences between the source- and target-conditioned outputs; cooler colours indicate closer agreement. TAR uses this agreement to weight source-residual injection at overlapping source and target coordinates.}}
\label{fig:tar-vis-1b}
\end{figure}

\clearpage
\subsection{Large-scale qualitative gallery}
\label{app:gallery}

To further demonstrate the generality and robustness of VS3D, we present an extended gallery of editing results across a wide variety of 3D assets and edit instructions. These examples span diverse local-edit operations as well as mixed-operation edits, covering diverse object categories including characters, animals, vehicles, and everyday objects. The results confirm that VS3D consistently produces high-fidelity edits while preserving non-edited regions, even under challenging compositional instructions.

\begin{figure}[H]
\centering
\includegraphics[width=.90\textwidth]{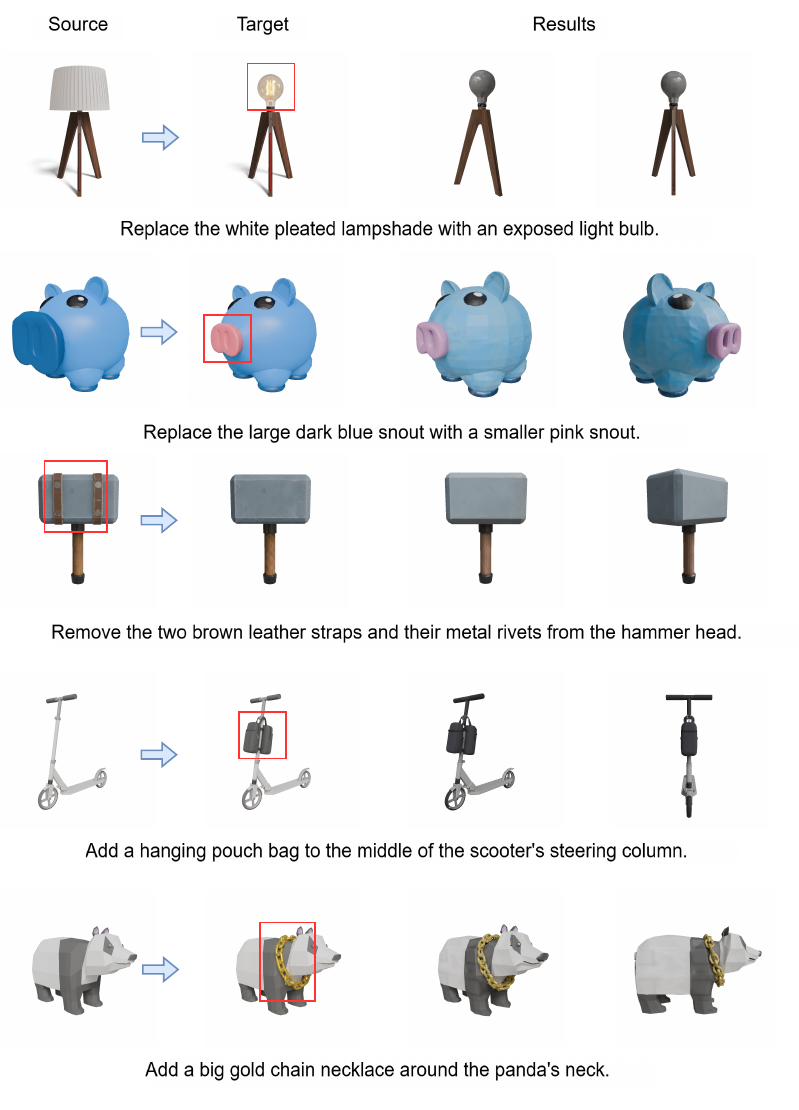}
\caption{\textbf{Extended editing gallery (Part 1).} Large-scale qualitative results of VS3D across diverse assets and edit operations. Each group shows the source asset and the edited result under the corresponding text instruction.}
\label{fig:gallery-1}
\end{figure}

\clearpage
\begin{figure}[H]
\centering
\includegraphics[width=\textwidth]{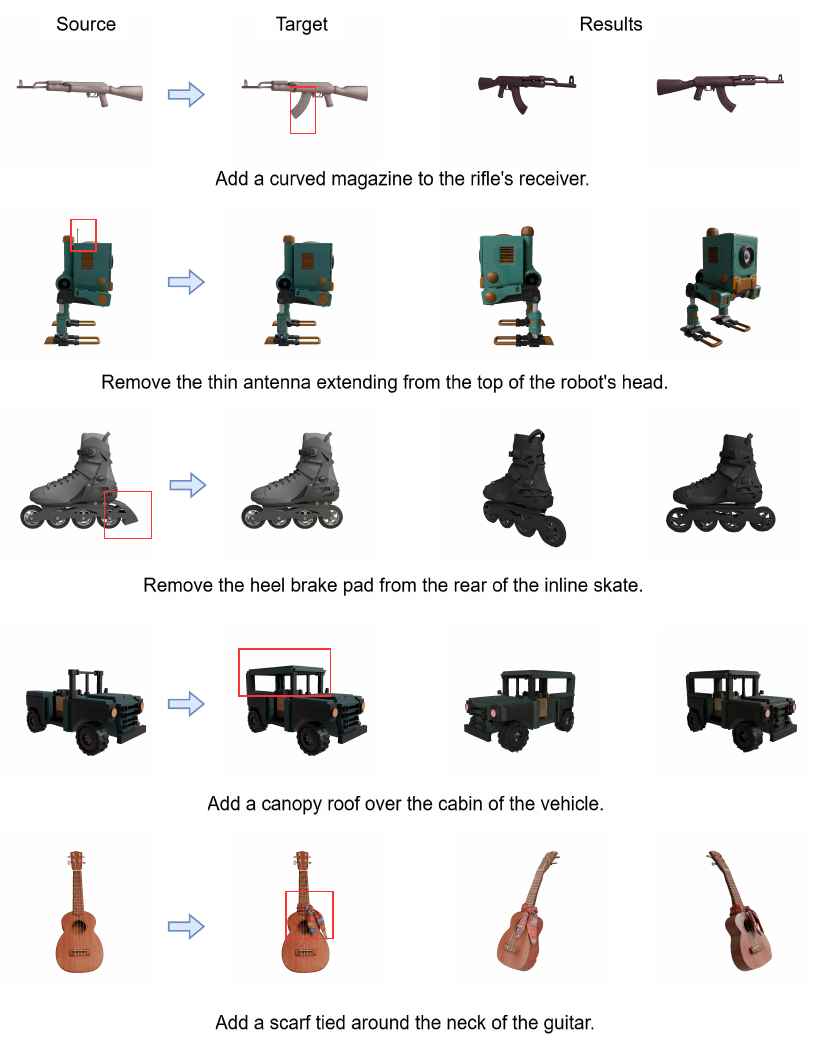}
\caption{\textbf{Extended editing gallery (Part 2).} Additional large-scale results demonstrating VS3D's robustness on mixed-operation edits and complex compositional instructions.}
\label{fig:gallery-2}
\end{figure}


\newpage
\section*{NeurIPS Paper Checklist}

\begin{enumerate}

\item {\bf Claims}
    \item[] Question: Do the main claims made in the abstract and introduction accurately reflect the paper's contributions and scope?
    \item[] Answer: \answerYes{}
    \item[] Justification: The abstract and introduction clearly state that VS3D is a training-free, mask-free 3D editing framework operating in velocity space, which is evaluated through the quantitative, qualitative, ablation, benchmark, and user-study experiments in Section~4 and Appendix~A.

\item {\bf Limitations}
    \item[] Question: Does the paper discuss the limitations of the work performed by the authors?
    \item[] Answer: \answerYes{}
    \item[] Justification: \textcolor{black}{Section~\ref{sec:discussion} discusses residual geometry and material changes in non-edited regions, the source-preserving editing scope, and dependence on the generative capacity of the TRELLIS~2.0 backbone.}

\item {\bf Theory assumptions and proofs}
    \item[] Question: For each theoretical result, does the paper provide the full set of assumptions and a complete (and correct) proof?
    \item[] Answer: \answerNA{}
    \item[] Justification: This paper does not include formal theoretical results or proofs; it is a methods paper validated empirically.

    \item {\bf Experimental result reproducibility}
    \item[] Question: Does the paper fully disclose all the information needed to reproduce the main experimental results of the paper to the extent that it affects the main claims and/or conclusions of the paper (regardless of whether the code and data are provided or not)?
    \item[] Answer: \answerYes{}
    \item[] Justification: \textcolor{black}{Algorithm~\ref{alg:full} describes the inference pipeline, and Appendix~\ref{app:setup} reports the benchmark construction, metrics, and implementation settings. Our method builds on the publicly available TRELLIS~2.0 model; implementation and setup instructions are available at \url{https://github.com/LHaoooo/VS3D}.}

\item {\bf Open access to data and code}
    \item[] Question: Does the paper provide open access to the data and code, with sufficient instructions to faithfully reproduce the main experimental results, as described in supplemental material?
    \item[] Answer: \answerNo{}
    \item[] Justification: \textcolor{black}{The implementation and setup instructions are publicly available at \url{https://github.com/LHaoooo/VS3D}. The benchmark data are not yet fully public and will be released progressively. Therefore, complete public access to the data needed to reproduce the reported benchmark results is not yet available.}

\item {\bf Experimental setting/details}
    \item[] Question: Does the paper specify all the training and test details (e.g., data splits, hyperparameters, how they were chosen, type of optimizer) necessary to understand the results?
    \item[] Answer: \answerYes{}
    \item[] Justification: \textcolor{black}{Section~\ref{sec:exp} describes the comparisons and ablations, and Appendix~\ref{app:setup} reports the benchmark construction, metrics, and implementation settings. The user-study protocol is provided in Appendix~\ref{app:userstudy}.}

\item {\bf Experiment statistical significance}
    \item[] Question: Does the paper report error bars suitably and correctly defined or other appropriate information about the statistical significance of the experiments?
    \item[] Answer: \answerNo{}
    \item[] Justification:  \textcolor{black}{We report aggregate benchmark metrics and the user-study vote counts and percentages. We do not report confidence intervals or statistical significance tests.}

\item {\bf Experiments compute resources}
    \item[] Question: For each experiment, does the paper provide sufficient information on the computer resources (type of compute workers, memory, time of execution) needed to reproduce the experiments?
    \item[] Answer: \textcolor{black}{\answerNo{}}
    \item[] Justification: \textcolor{black}{Appendix~\ref{app:runtime} reports per-asset runtime on a single NVIDIA RTX 4090 GPU for VS3D and four baselines. Peak memory usage and the total compute used across all experiments are not reported.}
    
\item {\bf Code of ethics}
    \item[] Question: Does the research conducted in the paper conform, in every respect, with the NeurIPS Code of Ethics \url{https://neurips.cc/public/EthicsGuidelines}?
    \item[] Answer:  \textcolor{black}{\answerYes{}}
    \item[] Justification:  \textcolor{black}{The research includes a voluntary preference study with 21 participants. Participants were informed about the study procedure before participating, and no compensation was provided.}

\item {\bf Broader impacts}
    \item[] Question: Does the paper discuss both potential positive societal impacts and negative societal impacts of the work performed?
    \item[] Answer: \textcolor{black}{\answerYes{}}
    \item[] Justification: \textcolor{black}{The broader-impact paragraph in Appendix~\ref{app:broader-impact} discusses benefits for creative workflows, potential unauthorized asset modification and misleading content, and responsible use respecting asset licenses and permissions.}
    
\item {\bf Safeguards}
    \item[] Question: Does the paper describe safeguards that have been put in place for responsible release of data or models that have a high risk for misuse (e.g., pre-trained language models, image generators, or scraped datasets)?
    \item[] Answer: \answerNA{}
    \item[] Justification: VS3D does not release new pre-trained models or scraped datasets; it operates on an existing publicly available model (TRELLIS~2.0).

\item {\bf Licenses for existing assets}
    \item[] Question: Are the creators or original owners of assets (e.g., code, data, models), used in the paper, properly credited and are the license and terms of use explicitly mentioned and properly respected?
    \item[] Answer: \answerYes{}
    \item[] Justification: TRELLIS~2.0 (MIT License) and Objaverse (ODC-By License) are properly cited. All 3D assets used are from publicly available datasets with permissive licenses.

\item {\bf New assets}
    \item[] Question: Are new assets introduced in the paper well documented and is the documentation provided alongside the assets?
    \item[] Answer: \textcolor{black}{\answerNo{}}
    \item[] Justification: \textcolor{black}{The paper introduces an Objaverse-derived editing benchmark and releases the VS3D implementation, but no new pretrained model weights. Appendix~\ref{app:setup} documents the benchmark construction. The benchmark data will be released progressively; a complete public benchmark package with accompanying documentation is not yet available.}

\item {\bf Crowdsourcing and research with human subjects}
    \item[] Question: For crowdsourcing experiments and research with human subjects, does the paper include the full text of instructions given to participants and screenshots, if applicable, as well as details about compensation (if any)? 
    \item[] Answer: \answerNo{}
    \item[] Justification:  \textcolor{black}{Appendix~\ref{app:userstudy} describes the study protocol and states that participants received no compensation. The original participant instructions and interface screenshots were not retained, so they cannot be provided.}

\item {\bf Institutional review board (IRB) approvals or equivalent for research with human subjects}
    \item[] Question: Does the paper describe potential risks incurred by study participants, whether such risks were disclosed to the subjects, and whether Institutional Review Board (IRB) approvals (or an equivalent approval/review based on the requirements of your country or institution) were obtained?
    \item[] Answer: \answerNo{}
    \item[] Justification:  \textcolor{black}{The study involved voluntary preference judgments on rendered 3D-editing results. Participants were informed of the study procedure beforehand and received no compensation. We anticipated no risks beyond viewing these outputs. The supervising faculty member agreed to the study being conducted; however, no formal IRB or equivalent institutional ethics review was obtained.}

\item {\bf Declaration of LLM usage}
    \item[] Question: Does the paper describe the usage of LLMs if it is an important, original, or non-standard component of the core methods in this research? Note that if the LLM is used only for writing, editing, or formatting purposes and does \emph{not} impact the core methodology, scientific rigor, or originality of the research, declaration is not required.
    \item[] Answer: \answerNA{}
    \item[] Justification: LLMs are not used as a component in the core methodology of VS3D.

\end{enumerate}

\end{document}